\documentclass[usenatbib]{mn2e}
\usepackage{amssymb,amsmath,graphicx}

\newcommand{\Jvec}{\textit{\textbf{J}} }
\newcommand{\wvec}{\textit{\textbf{w}} }
\newcommand{\zvec}{\textit{\textbf{z}} }
\newcommand{\xvec}{\textit{\textbf{x}} }
\newcommand{\vvec}{\textit{\textbf{v}} }
\newcommand{\kvec}{\textit{\textbf{k}} }
\newcommand{\Evec}{\textit{\textbf{E}} }
\newcommand{\Psivec}{{\bf \Psi}}
\newcommand{\avec}{\textit{\textbf{a}} }
\newcommand{\bvec}{\textit{\textbf{b}} }

\begin{document}

\title{Finite element modelling of perturbed stellar systems}
\author[Mir Abbas Jalali]
  {Mir Abbas~Jalali\thanks{mjalali@sharif.edu (MAJ)} \\
   Sharif University of Technology, Postal Code: 14588-89694, 
   Azadi Avenue, Tehran, Iran}
   
\maketitle

\begin{abstract}
I formulate a general finite element method (FEM) for self-gravitating 
stellar systems. I split the configuration space to finite elements, 
and express the potential and density functions over each element 
in terms of their nodal values and suitable interpolating functions. 
General expressions are then introduced for the Hamiltonian and phase 
space distribution functions of the stars that visit a given element. 
Using the weighted residual form of Poisson's equation, I derive the 
Galerkin projection of the perturbed collisionless Boltzmann equation, 
and assemble the global evolutionary equations of nodal distribution 
functions. The FEM is highly adaptable to all kinds of potential and 
density profiles, and it can deal with density clumps and initially 
non-axisymmetric systems. I use ring elements of non-uniform widths, 
choose linear and quadratic interpolation functions in the radial 
direction, and apply the FEM to the stability analysis of the cutout 
Mestel disc. I also integrate the forced evolutionary equations and 
investigate the disturbances of a stable stellar disc due to the 
gravitational field of a distant satellite galaxy. The performance 
of the FEM and its prospects are discussed. 
\end{abstract}


%\numberwithin{equation}{section}

\section{Introduction}

Schwarzschild's (1979) orbit superposition method, $N$-body simulations 
\citep{BT08} and smoothed particle hydrodynamics (SPH) \citep{SH02,Sp05}
have been the main simulation tools of dynamical processes in star clusters, 
galaxies and dark matter halos. As parallel computers are developed and 
special-purpose hardware cards emerge \citep{S90,M03,PZ07,PZ08,GHPZ09}, 
the resolution of simulations is enhanced too. Nevertheless, the number 
of particles that the most sophisticated codes and hardwares can handle, 
still lags realistic figures by several orders of magnitude. Combined 
with the problem of setting initial conditions for a given galaxy, this 
technological limitation leaves the ground open for alternative methods 
like a direct search for the solutions of the Boltzmann equation. 
One such an idea was introduced in \citet{J07}, where the variational 
weighted residual form of the collisionless Boltzmann equation (CBE) 
was used to study the modal structure of disc galaxies, and the CBE 
was projected onto a system of ordinary differential equations. 

The success of variational methods, however, depends on the 
potential--density basis sets used in the derivation of test and trial 
functions. For instance, after 15 to 20 terms, inappropriate test and 
trial functions may contribute more to the noise than building the 
perturbed density in clumpy regions or in cusps. Our choices of 
potential--density basis functions are indeed limited both for discs 
\citep{CB72,K76,Q92,Q93} and for three dimensional systems 
\citep{CB73,HO92,Z96,RJ09}, and it is not always possible to systematically 
find/tailor \citep{Saha91,RE96,W99} suitable potential--density pairs for a 
given galaxy model. We thus need a unified methodology with adaptively 
controlled resolution, and capable of modeling arbitrary density profiles. 

The existence of different kinds of orbits in a stellar system further 
complicates the governing dynamics. It is a routine procedure to work with 
initially axisymmetric disc galaxies because their phase space is filled 
by rosette orbits. In natural systems, however, small asymmetries generate 
higher-order resonant orbits, the angular momentum of individual orbits is 
no longer conserved and finding a physical phase space distribution 
function (DF) demands accurate knowledge about resonant tori. 
In such conditions, $N$-body and SPH simulations may lead to unrealistic results, 
especially near sharp clumps \citep{Agertz07} or at the boundaries of resonant
tori: the stars of thin tori may be totally missed in simulations or they may 
cause noise and artificial heating rather than participating in local/global 
structure formation. Schwarzschild's (1979) method can still be trusted when 
it comes to taking into account (theoretically) all possible orbit families, 
but it cannot be efficiently applied to the modeling of transient processes. 
In this study, I present a finite element method (FEM) for investigating the 
time-dependent evolution of stellar systems. Over half a century, 
the FEM has undoubtedly played a crucial role in structural engineering, 
geophysics and fluid dynamics \citep{ZT05,LNS04,parker08}, but its 
application to stellar dynamics is initiated here. 

After a brief review of orbit calculations on resonant tori, I formulate 
a finite element technique in the configuration space for solving Poisson's 
equation/integral, and describe the perturbed potential and density 
functions of a disc galaxy in the vicinity of a general non-axisymmetric 
equilibrium. I then express the distribution and Hamiltonian functions 
in terms of local angle-action variables on tori and over finite elements. 
I derive a relation between nodal densities and DFs, and use Galerkin's 
projection to obtain a system of ordinary differential equations for the 
temporal evolution of DF. I discuss the nature of evolutionary equations 
in the absence and presence of external perturbers, and validate my FEM 
code by investigating the linear stability problem of the stellar Mestel 
disc. I then study the disturbances induced by a satellite galaxy on its 
primary stellar disc. I conclude the paper by comparing the FEM with 
other techniques, and discuss about its possible developments.

\section{Finite element formulation}
\label{sec:FEM-formulation}

The evolution of a stellar system near a given equilibrium 
state is described by the phase space DF 
\begin{eqnarray}
f(\xvec,\vvec,t)=
f_0(\xvec,\vvec)+
f_1(\xvec,\vvec,t),
\end{eqnarray}
where $\xvec=(x,y,z)$ and $\vvec=(v_x,v_y,v_z)$ 
are, respectively, the Cartesian coordinates (of stars) and their 
conjugate momenta, and $t$ is the time. The subscripts 0 and 1 denote 
equilibrium and perturbed states, respectively. In a collisionless 
system the function $f$ is conserved along the orbits of stars and one has
\begin{eqnarray}
\frac{\partial f}{\partial t}+\left [ f,{\cal H} \right ]=0,
\label{eq:CBE-x-v}
\end{eqnarray}
where $[\cdots,\cdots]$ is the Poisson bracket taken over 
the $(\xvec,\vvec)$-space, and 
\begin{eqnarray}
{\cal H}(\xvec,\vvec,t)=
\frac 12 \vvec\cdot \vvec
+\Phi_0(\xvec)+\Phi_1(\xvec,t)
+\Phi_{\rm e}(\xvec,t),
\label{eq:Hamiltonian}
\end{eqnarray}
is the Hamiltonian function that governs the motion of stars
subject to the mean-field potential 
$\Phi_0(\xvec)+\Phi_1(\xvec,t)$
due to self-gravity, and a weak external field 
$\Phi_{\rm e}(\xvec,t)$. The equilibrium and 
perturbed self-gravitational potentials are related to the 
densities $\rho_j=\int f_j {\rm d}^3\vvec$ 
through Poisson's equation $\nabla ^2 \Phi_j=4\pi G\rho_j$
($j$=0,1) where $G$ is the universal constant of gravitation. 
For discs, this equation is replaced by Poisson's integral
\begin{eqnarray}
\Phi_j = -G \int 
\frac{\Sigma_j {\rm d}^2\xvec'}
       {|\xvec'-\xvec|},
 \label{eq:Poisson-2D}       
\end{eqnarray}
where $\Sigma_j =\int f_j{\rm d}^2\vvec$ is the surface 
density. I am interested in solving the CBE for time-varying systems, 
so the perturbed potential and density functions depend on $t$,
explicitly.

Regular orbits of the initial Hamiltonian system 
\begin{eqnarray}
{\cal H}_0(\xvec,\vvec)=
\frac 12 \vvec \cdot \vvec
+\Phi_0(\xvec),
\end{eqnarray}
lie on invariant tori, and the set of tori with common central 
periodic orbits constitute a resonant bundle. I focus on initially 
integrable systems whose resonant bundles are separated by the 
invariant manifolds of unstable periodic orbits. Non-integrability 
is usually associated with the destruction of invariant manifolds
and the occurrence of a layer of chaotic orbits. Resonant bundles 
associated with ${\cal H}_0$ can be constructed in terms of the 
action variables $\Jvec$ and their conjugate angles 
$\wvec$ \citep{MB90,KB94a,KB94b}, and the phase space 
coordinates of stars are determined from
\begin{eqnarray}
\xvec(t) \!\! &=& \!\!
\sum_{ \kvec } 
\textit{\textbf{X}}_{\kvec }(\Jvec)
e^{{\rm i}\kvec \cdot \wvec },~~
\wvec={\bf \Omega} t+\wvec_0,~~
{\rm i}=\sqrt{-1}, 
\label{eq:x-vs-actions-angles} \\
\vvec(t) \!\! &=& \!\!
\sum_{ \kvec } {\rm i} 
\left (\kvec \cdot {\bf \Omega} \right ) 
\textit{\textbf{X}}_{\kvec }(\Jvec)
e^{{\rm i}\kvec \cdot \wvec }.~~
\textit{\textbf{X}}^{*}_{\kvec }(\Jvec)=
\textit{\textbf{X}}_{-\kvec }(\Jvec).
\label{eq:v-vs-actions-angles}
\end{eqnarray}
Here $\kvec$ is a vector of integer numbers,
${\bf \Omega}=\partial {\cal H}_0/\partial \Jvec$ is 
the vector of orbital frequencies and asterisk stands for complex 
conjugation. Substituting from (\ref{eq:x-vs-actions-angles}) 
and (\ref{eq:v-vs-actions-angles}) into 
$f(\xvec,\vvec,t)$ and 
${\cal H}(\xvec,\vvec,t)$ 
leads to   
\begin{eqnarray}
f \!\! &=& \!\! 
f_0(\Jvec)+f_1(\wvec,\Jvec,t),
~~ f_1={\rm Re} \sum_{\kvec} 
\tilde f_{1,\kvec}(\Jvec,t) 
e^{{\rm i}\kvec\cdot \wvec}, 
\label{eq:DF-in-the-angle-action-space} \\
{\cal H} \!\! &=& \!\!
{\cal H}_0(\Jvec) +
\Phi_1(\wvec,\Jvec,t)+
\Phi_{\rm e}(\wvec,\Jvec,t),
\label{eq:Hamiltonian-in-the-angle-action-space}
\end{eqnarray}
so that 
\begin{eqnarray}
\Phi_1 \!\! &=& \!\! {\rm Re} \sum_{\kvec} 
\tilde h_{1,\kvec}(\Jvec,t) 
e^{{\rm i}\kvec\cdot \wvec}, 
\label{eq:self-gravity-potential-angle-action} \\
\Phi_{\rm e} \!\! &=& \!\! {\rm Re} \sum_{\kvec} 
\tilde h_{{\rm e},\kvec}(\Jvec,t) 
e^{{\rm i}\kvec\cdot \wvec}.
\label{eq:external-potential-angle-action}
\end{eqnarray}
The dependency of $f_0$ only on the action variables is deduced 
from Jeans theorem. I intend to determine the functions 
$\tilde f_{1,\kvec}(\Jvec,t)$ 
and $\tilde h_{1,\kvec}(\Jvec,t)$ over a set 
of finite elements. The advantage of using angle-action variables is 
that physical quantities are modelled in terms of Fourier series over 
half of the phase space. I continue with the finite element formulation 
of stellar discs and the generalisation of the same procedure to three 
dimensional systems will be presented elsewhere.

\subsection{Finite ring elements in the configuration space}
\label{subsec:Galerkin-projection-Possion}

Adopting the usual polar coordinates $(R,\phi)$, where $R$ is  
the radial distance from the galactic centre and $\phi$ is the 
azimuthal angle, the $2\pi$-periodicity of physical quantities 
in the $\phi$-direction suggests the Fourier expansions of the 
perturbed potential and surface density as
\begin{eqnarray}
\Phi_1(R,\phi,t) \!\! &=& \!\! {\rm Re} \sum_{m=-\infty}^{+\infty} 
P_m(R,t) e^{{\rm i} m\phi},
\label{eq:expand-pot}
\\
\Sigma_1(R,\phi,t) \!\! &=& \!\! {\rm Re} \sum_{m=-\infty}^{+\infty} 
S_m(R,t) e^{{\rm i} m\phi}.
\label{eq:expand-dens}
\end{eqnarray}
The configuration space is then split to $N$ ring elements. 
The width of the $n$th element is obtained using its nodal 
radii $R_n$ and $R_{n+1}$ as $\Delta R_n = R_{n+1}-R_n$, and 
the functions $P_m(R,t)$ and $S_m(R,t)$ are approximated by
\begin{eqnarray}
P_m(R,t) \!\! &=& \!\! \sum_{n=1}^{N} H_n(R)
\textit{\textbf{G}}_n \cdot \avec^n_m, 
\label{eq:expand-pot-elements} \\
S_m(R,t) \!\! &=& \!\! \sum_{n=1}^{N} H_n(R) 
\textit{\textbf{G}}_n \cdot \bvec^n_m, 
\label{eq:expand-dens-elements} \\
H_n(R) &=& \left \{ 
\begin{array}{ll}
1, & ~~ R_n \le R \le R_{n+1}, \\
0, & ~~ R<R_n ~~ {\rm or} ~~ R>R_{n+1}.
\end{array}
\right.
\label{eq:define-HnR}
\end{eqnarray}
The elements of the $N_{\rm d}$-dimensional row vector 
\begin{eqnarray}
\textit{\textbf{G}}_n=\left [ 
\begin{array}{llll}
G_{1n}(R) & G_{2n}(R) & \cdots & G_{N_{\rm d} n}(R)
\end{array}
\right ],
\end{eqnarray}
are suitable interpolating functions (also known as shape functions) 
in the $R$-domain, and the $N_{\rm d}$-dimensional column vectors 
$\avec^n_m$ and $\bvec^n_m$ are time-dependent nodal amplitudes 
defined by 
\begin{eqnarray}
\avec^n_m \!\!\! &=& \!\!\!
\left [ 
\begin{array}{llll} 
a^n_{1m}(t) & a^n_{2m}(t) & \cdots & a^n_{N_{\rm d} m}(t) 
\end{array}
\right ]^{\rm T}, \\
\bvec^n_m \!\!\! &=& \!\!\! 
\left [ 
\begin{array}{llll} 
b^n_{1m}(t) & b^n_{2m}(t) & \cdots & b^n_{N_{\rm d} m}(t) 
\end{array}
\right ]^{\rm T}.
\end{eqnarray}
Here a superscript T stands for transpose and a dot denotes matrix/vector 
multiplication. The parameter $N_{\rm d}$ is the number of nodes in a 
single element, and we have the general property $G_{jn}(\bar R_k)=\delta_{jk}$, 
where $\delta_{jk}$ is the Kronecker delta and $\bar R_k$ is the radial 
position of the $k$th node in the normalised coordinate:
\begin{eqnarray}
\bar R = 2\frac{R-R_n}{\Delta R_n}-1.
\end{eqnarray}

The simplest one dimensional element is obtained for $N_{\rm d}=2$ 
by using the linear functions 
\begin{eqnarray}
G_{1n} = \frac{1}{2}(1-\bar R),~~
G_{2n} = \frac{1}{2}(1+\bar R).
\label{eq:linear-interpolate}
\end{eqnarray}
For $N_{\rm d}>2$, higher-order elements with $N_{\rm d}-2$ interior 
nodes are built. The interpolating functions of a quadratic element 
of the so-called serendipity family are 
\begin{eqnarray}
G_{1n} = \frac{{\bar R}^2-\bar R}{2},~
G_{2n} = 1-{\bar R}^2,~
G_{3n} = \frac{\bar R+{\bar R}^2}{2}.
\label{eq:quadratic-interpolate}
\end{eqnarray}
These linear and quadratic functions are of $C_0$ class,
which guarantee that $S_m(R,t)$ and $P_m(R,t)$ are smooth 
(continuous and differentiable) inside elements and continuous at 
the boundary nodes should the amplitude functions satisfy
\begin{eqnarray}
a^n_{N_{\rm d} m}=a^{(n+1)}_{1m},~~b^n_{N_{\rm d} m}=b^{(n+1)}_{1m},
~~n\ge 1. \label{eq:continuity-condition}
\end{eqnarray}

On substituting from (\ref{eq:expand-pot}) and (\ref{eq:expand-dens}) 
into (\ref{eq:Poisson-2D}) and changing the integration variables 
to polar coordinates, one obtains
\begin{eqnarray}
P_m(R,t) \!\! &=& \!\! \lim _{\epsilon \rightarrow 0} \Bigg [
\frac{-2G}{\sqrt{R}} \int_{0}^{\infty} {\rm d}R'
\sqrt{R'} S_m(R',t) \nonumber \\
\!\! &\times & \!\! Q_{m-1/2} \left ( 
\frac{\epsilon^2+R^2+R'^2}{2RR'} \right ) \Bigg ],
\label{eq:pot-m-versus-dens-m}
\end{eqnarray}
where $Q_{\nu}(z)$ is the associated Legendre function of
the second kind. The parameter $\epsilon$ is introduced to 
handle the divergence of $Q_{\nu}(z)$ at $|z|=1$. 
although $Q_{m-1/2}(1)$ is indefinite, the limit of the 
whole bracketed statement in (\ref{eq:pot-m-versus-dens-m}) 
exists as $\epsilon \rightarrow 0$. I substitute the series
of (\ref{eq:expand-pot-elements}) and (\ref{eq:expand-dens-elements}) 
into (\ref{eq:pot-m-versus-dens-m}), take the inner product of 
the resulting equation by $H_{n'}(R)\textbf{\textit{G}}^{\rm T}_{n'}$
and carry out the integrations over $R$ and $R'$ to obtain 
\begin{eqnarray}
\avec^{n'}_{m}(t)=
-2G \sum_{n=1}^{N} \left [ \textsf{\textbf{A}}^{-1}(n') \cdot   
\textsf{\textbf{B}}(m,n',n) \right ] \cdot 
\bvec^{n}_{m}(t),
\label{eq:reduced-poisson}
\end{eqnarray}
for $n'=1,2,\cdots,N$. Equation (\ref{eq:reduced-poisson}) 
is the Galerkin projection (or weighted residual form) of 
Poisson's integral over the $n'$th element. The constant 
$N_{\rm d}\times N_{\rm d}$ matrix $\textsf{\textbf{A}}$ 
is defined as
\begin{eqnarray}
\textsf{\textbf{A}}(n)=
\int_{R_n}^{R_{n+1}} \textbf{\textit{G}}^{\rm T}_{n}(R) \cdot 
 \textbf{\textit{G}}_n(R) ~ {\rm d}R,
\end{eqnarray}
and the elements of the $N_{\rm d} \times N_{\rm d}$ matrix 
$\textsf{\textbf{B}}=[B_{ij}]$ are computed from 
\begin{eqnarray}
B_{ij}(m,n',n) \!\!\! &=& \!\!\! \lim _{\epsilon \rightarrow 0}
\int_{R_{n'}}^{R_{n'+1}}\int_{R_n}^{R_{n+1}} 
\sqrt{\frac{R'}{R}} G_{in'}(R) G_{jn}(R') \nonumber \\
\!\!\! &\times& \!\!\! Q_{m-1/2}\left ( \frac{\epsilon^2+R^2+R'^2}{2RR'} \right )
{\rm d}R' ~{\rm d}R. 
\end{eqnarray}
Considering the conditions in (\ref{eq:continuity-condition}), nodal 
potentials and densities can be collected, respectively, in the column 
vectors
\begin{eqnarray}
\textit{\textbf{p}}_m(t) \!\! &=& \!\! \left [
\begin{array}{llll}
p^1_{m}(t) & p^2_{m}(t) & \cdots & p^{N_{\rm t}}_m(t)
\end{array}
\right ]^{\rm T}, \\
\textit{\textbf{d}}_m(t) \!\! &=& \!\! \left [
\begin{array}{llll}
d^1_m(t) & d^2_m(t) & \cdots & d^{N_{\rm t}}_m(t)
\end{array}
\right ]^{\rm T},
\end{eqnarray}
with $N_{\rm t}$ being the total number of boundary and interior 
nodes. The system of $N_{\rm d} \times N$ linear equations 
(\ref{eq:reduced-poisson}) is thus assembled to
\begin{eqnarray}
\textit{\textbf{p}}_m(t)=
\textsf{\textbf{C}}(m) \cdot \textit{\textbf{d}}_m(t),
\label{eq:reduced-poisson-all}
\end{eqnarray}
where $\textsf{\textbf{C}}(m)$ is a generally dense 
$N_{\rm t} \times N_{\rm t}$ constant matrix. Equation 
(\ref{eq:reduced-poisson-all}) relates a discrete set 
of densities to their corresponding potentials. 

As an example, I use the FEM and construct the potential functions 
associated with Clutton-Brock's (1972) density functions
\begin{eqnarray}
\sigma^{m}_j(R) = \frac{2m+2j+1}{2\pi b^2}
\left ( \! \frac{b^2}{R^2+b^2} \! \right )^{3/2} \!
P^{m}_i\left ( \! \frac{R^2-b^2}{R^2+b^2} \! \right ),
\label{eq:dens-CB}
\end{eqnarray}
where $b$ is a length scale and $P^{m}_i$ are associated
Legendre functions with $i=m+j$. The functions $\sigma^{m}_j$ 
are oscillatory versus $R$ when $j\ge 1$. Figure 1a displays 
$\sigma^{2}_2(R)$ for $b=1$. I use linear interpolation functions 
with $N_{\rm d}=2$, and divide the $R$-domain to $N=25$ ring elements 
whose nodal radii are determined using the rule 
(there are no interior nodes)
\begin{eqnarray}
R_n = -\alpha_1 \ln \left (1- \frac{1}{2(N+1)} -\frac {n-1}{N+1} \right ).
\label{eq:nodal-radii}
\end{eqnarray}
The width $\Delta R_n$ of elements increases as one departs from the 
galactic centre. The parameter $\alpha_1$ determines the concentration 
of elements near the centre (or at large radii). Filled circles in 
Figure 1a mark the nodal densities $\sigma^2_2(R_n)$ ($n=1,2,\cdots,N+1$) 
that have been computed from (\ref{eq:dens-CB}) for $\alpha_1=2$. 
There is no time-dependence for a static mass distribution. 
The corresponding nodal potentials are therefore computed through
solving the linear system (\ref{eq:reduced-poisson-all}) for 
$\textit{\textbf{p}}_m$. Figure 1b shows the potential $\psi^2_2(R)$ 
calculated using Clutton-Brock's exact formula (solid line) and the 
FEM (scattered triangles). The agreement between 
analytical and finite element solutions is impressive: while the 
maximum magnitude of $|\psi^2_2(R)|$ is $0.674$, the root 
mean squared error  
\begin{eqnarray}
E_{\rm rms}=\left \{ \frac {1}{N+1} \sum_{j=1}^{N+1} 
\left [ p^j_{m}-\psi^2_2(R_j) \right ]^2 \right \}^{1/2},
\end{eqnarray}
is $E_{\rm rms}=0.0035$ for $N=25$ and it drops to $E_{\rm rms}=0.00097$ 
for $N=50$. Although the results are improved by further increasing 
the number of elements, switching to quadratic elements is more effective.
I added an interior node to the elements described in (\ref{eq:nodal-radii}) 
and used (\ref{eq:quadratic-interpolate}) with $N_{\rm d}=3$ to compute 
the potential function $\psi^2_2(R)$. For $N=50$ elements, I found 
$E_{\rm rms}=0.000296$, which is lower by a factor of $\approx 3.3$ 
than the error corresponding to the same number of linear elements.  

\begin{figure}
\centerline{\hbox{\includegraphics[width=0.48\textwidth]{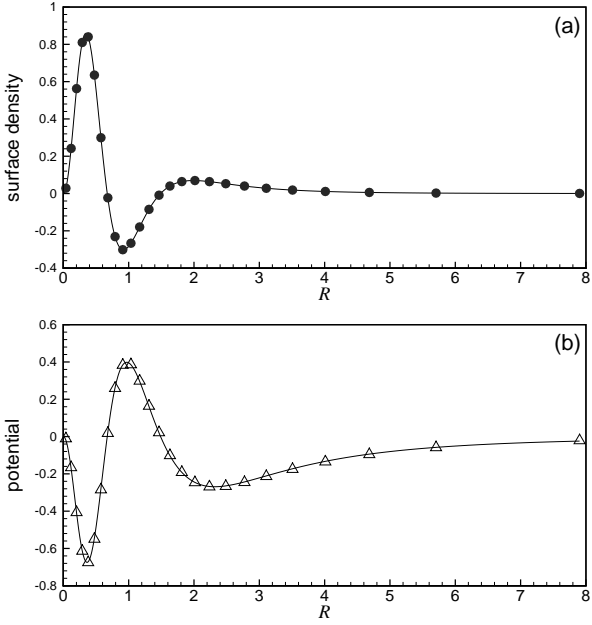}} }
\caption{(a) The graph of $\sigma^2_2(R)$ (solid line) and 
the nodal densities (filled circles) used in equation (\ref{eq:reduced-poisson-all}). 
(b) Exact analytical potential $\psi^2_2(R)$ (solid line) associated 
with $\sigma^2_2(R)$ and the finite element solution $\textit{\textbf{p}}_m$ 
(triangles) for $N=25$.} 
\label{fig1}
\end{figure}

\subsection{Transformations to the angle-action space}

To determine $\tilde f_{1,\kvec}(\Jvec,t)$, 
I assume
\begin{eqnarray}
\tilde f_{1,\kvec}(\Jvec,t) = 
\sum_{n=1}^{N} \Evec_{\kvec}(n,\Jvec) 
\cdot \zvec^n_{\kvec }(t),
\label{eq:expand-DF-elements}
\end{eqnarray}
where the elements of the $N_{\rm d}$-dimensional row vectors
\begin{eqnarray}
\Evec_{\kvec}(n,\Jvec)=\left [ \!\! 
\begin{array}{llll}
E_{1,\kvec}(n,\Jvec) & E_{2,\kvec}(n,\Jvec) & \!\! \cdots \!\! & 
E_{N_{\rm d},\kvec}(n,\Jvec)
\end{array}
\!\! \right ],
\end{eqnarray}
are interpolation functions in the $\Jvec$-space and the 
time-dependent $N_{\rm d}$-dimensional column vectors 
\begin{eqnarray}
\zvec^n_{\kvec}(t)=\left [ 
\begin{array}{llll}
\zvec^n_{1,\kvec}(t) & \zvec^n_{2,\kvec}(t) & \!\! \cdots \!\! & 
\zvec^n_{N_{\rm d},\kvec}(t)
\end{array}
\right ]^{\rm T},
\end{eqnarray}
are the nodal DFs. In other words, the function
\begin{eqnarray}
\hat f_n(\wvec,\Jvec,t)= {\rm Re}
\sum_{\kvec } e^{{\rm i}\kvec\cdot \wvec}
\Evec_{\kvec}(n,\Jvec) 
\cdot \zvec^n_{\kvec }(t),
\label{eq:perturbed-DF-nth-element}
\end{eqnarray}
is the perturbed distribution function of those stars that 
enter (at least once) to the $n$th element. Stars on highly 
elongated orbits may visit more than one element. This justifies
the summation over $n$ in (\ref{eq:expand-DF-elements}). If an orbit 
crosses, tangentially or transversally, the boundary of the $n$th 
and $(n+1)$th elements when its phase space coordinates are 
$(\wvec_b,\Jvec_b)$, the condition $\hat f_n(\wvec_b,\Jvec_b,t)$=
$\hat f_{n+1}(\wvec_b,\Jvec_b,t)$ must be fulfilled, which gives 
\begin{eqnarray}
\Evec_{\kvec}(n,\Jvec_b) 
\cdot \zvec^n_{\kvec }(t) =
\Evec_{\kvec}(n+1,\Jvec_b) 
\cdot \zvec^{n+1}_{\kvec }(t).
\label{eq:constraint-Ek-boundary}
\end{eqnarray}
To ease this constraint on the elements of $\zvec^n_{\kvec}(t)$, 
one can choose the interpolating vector $\Evec_{\kvec}(n,\Jvec)$
so that 
\begin{eqnarray}
&{}& E_{N_{\rm d},\kvec}(n,\Jvec_b) = E_{1,\kvec}(n+1,\Jvec_b), 
\label{eq:equate-nodal-Ek} \\
&{}& \left \{
\begin{array}{ll}
E_{j,\kvec}(n,\Jvec_b) = 0, ~ & j<N_{\rm d}, \\
E_{j,\kvec}(n+1,\Jvec_b) = 0, ~ & j>1,
\end{array}
\right.
\label{eq:annul-nodal-Ek}
\end{eqnarray}
and reduce (\ref{eq:constraint-Ek-boundary}) to
\begin{eqnarray}
z^n_{N_{\rm d},\kvec}(t) = z^{n+1}_{1,\kvec}(t).
\label{eq:continuity-for-zk}
\end{eqnarray}
Equations (\ref{eq:equate-nodal-Ek})--(\ref{eq:continuity-for-zk}) 
constitute the continuity conditions of the perturbed DF over the 
ensemble of orbits that migrate between finite elements in the 
$\xvec$-space. In \S\ref{sec:find-Ek-vector}, I will introduce a 
method to find $\Evec_{\kvec}(n,\Jvec)$.
 
In order to satisfy $\Sigma_j =\int f_j{\rm d}^2\vvec$, one needs to 
establish a relation between the functions $\zvec^n_{\kvec }(t)$ and 
the nodal densities $\bvec^n_m(t)$. For doing so, I substitute from 
(\ref{eq:DF-in-the-angle-action-space}) and (\ref{eq:expand-dens}) 
into the fundamental equation 
\begin{eqnarray}
\Sigma_1(R,\phi,t) R ~ {\rm d}R ~{\rm d}\phi
=f_1(\wvec,\Jvec,t) 
~ {\rm d}^2 \Jvec ~{\rm d}^2 \wvec,
\end{eqnarray}
and use equations (\ref{eq:expand-dens-elements}) and 
(\ref{eq:expand-DF-elements}) to obtain
\begin{eqnarray}
&{}& \!\!\! \sum_{m=-\infty}^{+\infty} \sum_{n=1}^{N} H_n(R) 
\textit{\textbf{G}}_n \cdot \bvec^n_m(t) e^{{\rm i} m\phi}
R~{\rm d}R~{\rm d}\phi = \nonumber \\
&{}& \!\!\! \sum_{\kvec}
\sum_{n=1}^{N} \Evec_{\kvec}(n,\Jvec) 
\cdot \zvec^n_{\kvec }(t) 
e^{{\rm i}\kvec\cdot \wvec} 
{\rm d}^2 \Jvec ~{\rm d}^2 \wvec. 
\label{eq:fundamental-DF-dens}
\end{eqnarray}
Taking the inner product of (\ref{eq:fundamental-DF-dens}) by 
$H_{n'}(R)\textbf{\textit{G}}^{\rm T}_{n'}\exp(-{\rm i}m'\phi)$, 
and carrying out the integrations over the $(R,\phi)$ and 
$(\Jvec,\wvec)$ spaces, result in 
the weighted residual form of the fundamental equation as
\begin{eqnarray}
\!\!\! &{}& \!\!\! \textsf{\textbf{K}}(n') \cdot 
\bvec^{n'}_{m'}(t) = \sum_{\kvec} 
\sum_{n=1}^{N} H_{n'}(R) \nonumber \\
\!\!\!\! &\times& \!\!\!\! \int \!\! \int e^{-{\rm i} m'\phi} 
\left [ \textbf{\textit{G}}^{\rm T}_{n'} \cdot 
\Evec_{\kvec}(n,\Jvec) \right ]
\cdot \zvec^n_{\kvec }(t)
e^{{\rm i}\kvec\cdot \wvec}
{\rm d}^2 \Jvec ~{\rm d}^2 \wvec,
\label{eq:fundamental-DF-dens-expanded}
\end{eqnarray}
where 
\begin{eqnarray}
\textsf{\textbf{K}}(n) = 2\pi \int_{R_n}^{R_{n+1}}
\textbf{\textit{G}}^{\rm T}_{n}(R) \cdot 
\textbf{\textit{G}}_{n}(R) ~
 R ~ {\rm d}R.
\end{eqnarray}

According to (\ref{eq:x-vs-actions-angles}), the radial distance
$R(t)=\sqrt{\xvec(t) \cdot \xvec(t) }$ 
and $\exp[{\rm i}\phi(t)]=[x(t)+{\rm i}y(t)]/R(t)$ admit Fourier 
series in terms of $\wvec$ so does the function
\begin{eqnarray}
H_{n'}(R)\textbf{\textit{G}}_{n'} e^{-{\rm i}m'\phi}=
\sum_{\kvec' } 
\Psivec_{\kvec' }(m',n',\Jvec)
e^{-{\rm i}\kvec' \cdot \wvec},
\label{eq:interpolating-fncs-fourier-expansion}
\end{eqnarray}
where the $N_{\rm d}\times 1$ row vector $\Psivec_{\kvec}$
is determined from
\begin{eqnarray}
\Psivec_{\kvec}(m',n',\Jvec) = \frac{1}{(2\pi)^2} \int 
H_{n'}(R)\textbf{\textit{G}}_{n'} e^{-{\rm i}m'\phi}
e^{{\rm i}\kvec \cdot \wvec} ~{\rm d}^2\wvec.
\label{eq:fourier-coefficients}
\end{eqnarray}
The dependency of the integrand on $H_{n'}(R)$ shows that 
$\Psivec_{\kvec}(m',n',\Jvec)$ is non-zero only for orbits 
that visit the $n$th ring element. 
Our physical sense of (\ref{eq:interpolating-fncs-fourier-expansion}) 
is sharpened by summing up its components and setting $m=0$ 
to obtain
\begin{eqnarray}
H_{n}(R) \sum_{j=1}^{N_{\rm d}} G_{jn}(R) =
\sum_{\kvec} \sum_{j=1}^{N_{\rm d}}
\Psi_{j,\kvec}(0,n,\Jvec) ~e^{-{\rm i}\kvec \cdot \wvec}.
\label{eq:transit-time-fourier-expansion}
\end{eqnarray}
From the properties of interpolating functions in the 
$R$-domain one can deduce that the summation on the left 
hand side of (\ref{eq:transit-time-fourier-expansion}) 
is unity. This result and the time-averages theorem 
\citep{BT08} imply that the quantity
\begin{eqnarray}
\sum_{j=1}^{N_{\rm d}} \Psi_{j,{\bf 0}}(0,n,\Jvec) =
\frac {1}{(2\pi)^2} \int H_n(R) ~{\rm d}^2 \wvec,
\label{eq:time-fraction}
\end{eqnarray}
is the fraction of time that a star of action vector 
$\Jvec$ spends inside the $n$th element. 

Defining 
\begin{eqnarray}
\textsf{\textbf{D}}(\kvec,m',n',n)=\int
\Psivec^{\rm T}_{\kvec }(m',n',\Jvec) \cdot
 \Evec_{\kvec }(n,\Jvec) ~ {\rm d}^2 \Jvec,
\end{eqnarray}
and substituting from (\ref{eq:interpolating-fncs-fourier-expansion}) 
into (\ref{eq:fundamental-DF-dens-expanded}) lead to 
\begin{eqnarray}
\bvec^{n'}_{m'}(t) = 4\pi^2 \sum_{\kvec } 
\sum_{n=1}^{N} \left [ \textsf{\textbf{K}}^{-1}(n') \cdot 
\textsf{\textbf{D}}(\kvec,m',n',n) \right ]
\cdot \zvec^n_{\kvec }(t),
\label{eq:nodal-dens-DF-relations}
\end{eqnarray}
which determines the vector of nodal densities in terms of 
$\zvec^n_{\kvec }(t)$. Taking into account the constraints 
(\ref{eq:continuity-for-zk}) and collecting the components 
of $\zvec^n_{\kvec }(t)$ (for $n=1,2,\cdots,N$) in a single 
vector $\zvec_{\kvec }(t)$, and combining the system of linear 
equations (\ref{eq:nodal-dens-DF-relations}) for all ring 
elements in the configuration space, result in
\begin{eqnarray}
\textit{\textbf{d}}_{m}(t) = \sum_{\kvec } 
\textsf{\textbf{F}}({\kvec},m)
\cdot \zvec_{\kvec }(t),
\label{eq:nodal-dens-DF-relations-all}
\end{eqnarray}
where $\textsf{\textbf{F}}({\kvec},m)$ is an $N_{\rm t}\times N_{\rm t}$ 
square matrix. Let me define $\textsf{\textbf{L}}(\kvec,m)=\textsf{\textbf{C}}(m)
\cdot \textsf{\textbf{F}}({\kvec},m)$. Equations (\ref{eq:reduced-poisson-all}) 
and (\ref{eq:nodal-dens-DF-relations-all}) will self-consistently give 
the nodal potentials $\textit{\textbf{p}}_m(t)$ in terms of 
$\zvec_{\kvec }(t)$:
\begin{eqnarray}
\textit{\textbf{p}}_{m}(t) =  
\sum_{\kvec } \textsf{\textbf{L}}({\kvec},m)
\cdot \zvec_{\kvec }(t).
\label{eq:nodal-pots-DF-relations-all}
\end{eqnarray}

By equating (\ref{eq:expand-pot}) and (\ref{eq:self-gravity-potential-angle-action}),
and applying (\ref{eq:expand-pot-elements}) and 
(\ref{eq:interpolating-fncs-fourier-expansion}), I conclude that 
\begin{eqnarray}
\tilde h_{1,\kvec}(\Jvec,t)=
\sum_{m=-\infty}^{+\infty} \sum_{n=1}^{N} 
\Psivec_{-\kvec }(-m,n,\Jvec) 
\cdot \avec^n_m(t).
\label{eq:equating-perturbed-Hamiltonians}
\end{eqnarray}
This means that a star of the angle-action coordinates 
$(\wvec,\Jvec)$ will experience the perturbed self-gravitational 
potential field 
\begin{eqnarray}
\hat h_n(\wvec,\Jvec,t)= {\rm Re}
\sum_{m=-\infty}^{+\infty} \sum_{\kvec } 
e^{{\rm i}\kvec \cdot \wvec}
\Psivec_{-\kvec }(-m,n,\Jvec) 
\cdot \avec^n_m(t),
\end{eqnarray}
during its passage through the $n$th element. Consequently, 
the perturbed Hamiltonian associated with the stars that 
enter to the $n$th element becomes
\begin{eqnarray}
\hat {\cal H}_n(\wvec,\Jvec,t) 
= {\rm Re} \left [ \hat h_n(\wvec,\Jvec,t)
+ \sum_{\kvec } 
\tilde h_{{\rm e},\kvec }(\Jvec,t)
e^{{\rm i}\kvec \cdot \wvec} \right ].
\label{eq:perturbed-potential-series}
\end{eqnarray}

\subsection{Galerkin weighting of the linearised CBE}
\label{sec:galerkin-projection-CBE}

Given the functions $\hat f_n$, $\hat h_n$ and $\Phi_{\rm e}$
in the angle-action space, the CBE reads  
\begin{eqnarray}
&{}& 
\sum_{n=1}^{N} \left \{ \frac{\partial \hat f_n}{\partial t}
\! + \! \left [ f_0,\hat h_n \right ] \! + \! 
\left [ \hat f_n,{\cal H}_0 \right ] \right \}
\! + \! \left [ f_0,\Phi_{\rm e} \right ] = 
\nonumber \\
&{}& -\sum_{n=1}^{N} 
\left [ \hat f_n,\Phi_{\rm e} \right ]
\! - \! \sum_{n,n'=1}^{N} 
\left [ \hat f_n,\hat h_{n'} \right ].
\label{eq:expand-poisson-brackets}
\end{eqnarray}
In the present analysis, the higher-order terms $[ \hat f_n,\hat h_{n'}]$ 
and $[ \hat f_n,\Phi_{\rm e}]$ are ignored. On substituting from 
(\ref{eq:perturbed-DF-nth-element}) and (\ref{eq:perturbed-potential-series}) 
into (\ref{eq:expand-poisson-brackets}), taking the inner product of the 
resulting equation with 
$\exp( -{\rm i}\kvec' \cdot \wvec ) \Evec^{\rm T}_{\kvec'}(n',\Jvec)$
and integrating over $\wvec$ and $\Jvec$, the following system of ordinary 
differential equations 
\begin{eqnarray}
\!\!\! &{}& \!\!\! {\rm i} \sum_{n=1}^{\infty} \textsf{\textbf{E}}_1(n',n,\kvec') 
\cdot \frac{{\rm d}}{{\rm d}t} \zvec^{n}_{\kvec' }(t) =  
\sum_{n=1}^{\infty} \textsf{\textbf{E}}_2(n',n,\kvec') 
\cdot \zvec^{n}_{\kvec' }(t)  \nonumber \\
\!\!\! &{}& \!\!\! - \textit{\textbf{Z}}_{n'}(\kvec',t) 
- \sum_{n=1}^{\infty} 
\sum_{m=-\infty}^{+\infty} 
\textsf{\textbf{E}}_3(m,n',n,\kvec')\cdot 
\avec^{n}_{m}(t), 
\label{eq:perturbed-CBE-projected}
\end{eqnarray}
are obtained for $n'=1,2,\cdots,N$ so that
\begin{eqnarray}
\textsf{\textbf{E}}_1 \!\!\! &=& \!\!\! 
\int \Evec^{\rm T}_{\kvec }(n',\Jvec)
\cdot \Evec_{\kvec }(n,\Jvec)
~{\rm d}^2 \Jvec, \label{eq:define-E1} \\
\textsf{\textbf{E}}_2 \!\!\! &=& \!\!\! 
\int \left ( \kvec \cdot {\bf \Omega} \right )
\Evec^{\rm T}_{\kvec }(n',\Jvec)
\cdot \Evec_{\kvec }(n,\Jvec)
~{\rm d}^2 \Jvec, \label{eq:define-E2} \\
\textsf{\textbf{E}}_3 \!\!\! &=& \!\!\! 
\int \left ( \kvec \cdot 
\frac{\partial f_0}{\partial \Jvec} \right )
\Evec^{\rm T}_{\kvec }(n',\Jvec)
\cdot \Psivec_{-\kvec }(-m,n,\Jvec)
~{\rm d}^2 \Jvec, \label{eq:define-E3} \\
\textit{\textbf{Z}}_{n'} \!\!\! &=& \!\!\! 
\int \left ( \kvec \cdot 
\frac{\partial f_0}{\partial \Jvec} \right )
\Evec^{\rm T}_{\kvec }(n',\Jvec)
\tilde h_{{\rm e},\kvec }(\Jvec,t)
~{\rm d}^2 \Jvec.
\label{eq:define-Zn}
\end{eqnarray}
The matrices $\textsf{\textbf{E}}_1$, $\textsf{\textbf{E}}_2$ 
and $\textsf{\textbf{E}}_3$ have the dimension $N_{\rm d} \times N_{\rm d}$,
and the forcing term $\textit{\textbf{Z}}_{n'}(\kvec,t)$ 
is a column vector of dimension $N_{\rm d}$. The integrals 
in (\ref{eq:define-E1})--(\ref{eq:define-Zn}) are performed 
over a $\Jvec$-subspace whose orbits visit both the $n'$th and 
$n$th elements. The summation over $n$ in (\ref{eq:perturbed-CBE-projected}) 
shows a coupling between adjacent and also unconnected distant 
elements, which communicate their dynamical information through 
elongated orbits. For example, if two peaks of a density wave
lie on a given trajectory, deformation of that trajectory near 
one peak will influence its behaviour near the other one. 
Long-range interactions of this kind will be negligible in 
models populated by near-circular orbits, or when density  
perturbations rise and fall in harmony with the equilibrium 
density profile. In such conditions, one can keep in 
(\ref{eq:perturbed-CBE-projected}) only the terms of $n=n'$ 
and write
\begin{eqnarray}
\!\!\! &{}& \!\!\! {\rm i} \textsf{\textbf{E}}_1(n,\kvec') 
\cdot \frac{{\rm d}}{{\rm d}t} \zvec^{n}_{\kvec' }(t) =  
\textsf{\textbf{E}}_2(n,\kvec') \cdot \zvec^{n}_{\kvec' }(t)  
\nonumber \\
\!\!\! &{}& \!\!\! - \textit{\textbf{Z}}_{n}(\kvec',t) 
- \sum_{m=-\infty}^{+\infty} 
\textsf{\textbf{E}}_3(m,n,\kvec') \cdot 
\avec^{n}_{m}(t).
\label{eq:perturbed-CBE-projected-nth-element}
\end{eqnarray}
This is indeed the weighted residual form of the linearised 
CBE over the $n$th element:
\begin{eqnarray}
\frac{\partial \hat f_n}{\partial t}
\! + \! \left [ f_0,\hat {\cal H}_n \right ] \! + \! 
\left [ \hat f_n,{\cal H}_0 \right ] = 0,
\label{eq:linearized-CBE-nth-element}
\end{eqnarray}
with Poisson brackets taken over a phase subspace whose 
stars visit the $n$th element. The set of element CBEs are 
sufficient (but not necessary) conditions for the global 
equation (\ref{eq:expand-poisson-brackets}) to be satisfied.
I recommend the application of the Galerkin form 
(\ref{eq:perturbed-CBE-projected-nth-element}) to the 
modelling of minimally peaked waves like the one displayed 
in the top-right panel of Figure \ref{fig2}. 

Dropping the prime sign for brevity, and assembling either 
the system of equations (\ref{eq:perturbed-CBE-projected}) or 
(\ref{eq:perturbed-CBE-projected-nth-element}) result in 
\begin{eqnarray}
\textsf{\textbf{U}}_1(\kvec)  \cdot 
\frac{{\rm d}\zvec_{\kvec }(t)}{{\rm d}t} 
\!\!\! &=& \!\!\! -{\rm i} \textsf{\textbf{U}}_2(\kvec) \cdot 
\zvec_{\kvec }(t) + 
{\rm i} \textit{\textbf{Z}}(\kvec,t) \nonumber \\
\!\!\! &+& \!\!\! 
\sum_{m=-\infty}^{+\infty}  {\rm i}
\textsf{\textbf{U}}_3(\kvec,m) \cdot 
\textit{\textbf{p}}_{m}(t), 
\label{eq:perturbed-CBE-projected-assembled}
\end{eqnarray}
where $\textsf{\textbf{U}}_1$, $\textsf{\textbf{U}}_2$ 
and $\textsf{\textbf{U}}_3$ are $N_{\rm t}\times N_{\rm t}$ 
square matrices, the forcing function $\textit{\textbf{Z}}(\kvec,t)$ 
is a column vector of dimension $N_{\rm t}$, and the vector 
of nodal potentials $\textit{\textbf{p}}_{m}(t)$ is computed 
from (\ref{eq:nodal-pots-DF-relations-all}). The continuity 
conditions (\ref{eq:continuity-for-zk}) must be taken into 
account in the assembly process. It is therefore a standard 
procedure to integrate the linear system 
(\ref{eq:perturbed-CBE-projected-assembled}) over the time 
domain and monitor the evolution of $\tilde f_{1,\kvec}(\Jvec,t)$. 

In the absence of external perturbations, equation
(\ref{eq:perturbed-CBE-projected-assembled}) admits a solution 
of the form 
$\zvec_{\kvec }(t)=\exp(-{\rm i}\omega t)
{\bf \zeta}_{\kvec }$ and it is reduced to the following 
linear eigensystem:
\begin{eqnarray}
\sum_{m=-\infty}^{+\infty}\sum_{\kvec' }
\left [ \textsf{\textbf{U}}_3(\kvec,m) \cdot
 \textsf{\textbf{L}}(\kvec',m) \right ]
\cdot {\bf \zeta}_{\kvec' }
\!\!\! &=& \!\!\! \nonumber \\
\left [\textsf{\textbf{U}}_2(\kvec) 
-\omega \textsf{\textbf{U}}_1(\kvec) \right ]
\cdot {\bf \zeta}_{\kvec }~, &{}&
\label{eq:linear-eigensystem}
\end{eqnarray}
that can be solved for the eigenfrequency $\omega$ and 
its corresponding eigenvector (mode shape) 
${\bf \zeta}_{\kvec }$ using standard numerical 
packages. The general form of the eigensystem presented in 
(\ref{eq:linear-eigensystem}) applies to initially non-axisymmetric 
discs that host a rich family of orbits living on resonant bundles. 
In the limit of initially round systems, only rosette orbits 
fill the phase space and it is convenient to define the generalised 
momenta of a test star by $(p_1,p_2)=(\dot R,R^2\dot \phi)$. 
The vector of action variables $\Jvec=(J_1,J_2)$ 
is then calculated from the integrals 
\begin{eqnarray}
J_1=\frac{1}{2\pi} \oint p_1 {\rm d}R,~~
J_2=\frac{1}{2\pi} \oint p_2 {\rm d}\phi,
\end{eqnarray}
along the rosette orbits of angular momentum $L=p_2$ and energy 
$E$ so that 
\begin{eqnarray}
E\equiv {\cal H}_0(\Jvec) =
\frac{1}{2} \left ( p_1^2 + \frac{p_2^2}{R^2} \right )+\Phi_0(R).
\end{eqnarray}
The angle variables $\wvec=(w_1,w_2)$ conjugate to 
the actions $(J_1,J_2)$ evolve according to the linear law 
$w_i=\Omega_i t+w_i(0)$ ($i$=1,2) with 
$\Omega_i=\partial {\cal H}_0/{\partial J_i}$. The axisymmetry 
condition implies $\Psivec_{-\kvec }(-m,n,\Jvec) 
=\Psivec_{\kvec }(m,n,\Jvec)$ and 
\begin{eqnarray}
\Psivec_{(k_1,k_2) }(m,n,\Jvec) =0,
~~{\rm if}~~k_2\not = m.
\label{eq:psi-for-axisymmetric}
\end{eqnarray}
Consequently, the summation over $m$ is dropped in 
(\ref{eq:perturbed-CBE-projected-assembled})--(\ref{eq:linear-eigensystem}), 
and the governing equations are decoupled for different values of the 
wavenumber $m$. The spectrum of the eigenfrequency $\omega$ contains 
pure oscillatory (van Kampen) modes as well as unstable ones. 
Spurious eigenfrequencies may also appear in the spectrum due to 
finite element discretisation errors. Such cases can be rejected 
by investigating their corresponding mode shapes.

\subsection{Interpolating functions in the action space}
\label{sec:find-Ek-vector}

The choice of the row vector $\Evec_{\kvec }(n,\Jvec)$ has a remarkable 
effect on the performance of the FEM. A somewhat trivial approach is to 
identify the sub-domain of $\Jvec$-space whose orbits visit the $n$th 
element, and generate a finite element mesh in that sub-domain by setting 
$E_{i,\kvec }(n,\Jvec)$ to elementary two dimensional counterparts of the 
functions used in \S\ref{subsec:Galerkin-projection-Possion}. This procedure, 
however, is not computationally efficient because it increases the sizes of 
global matrices $\textsf{\textbf{L}}$ and $\textsf{\textbf{U}}_i$ ($i=1,2,3$). 
Below, I follow an alternative approach and show that the neutral solutions 
of (\ref{eq:linearized-CBE-nth-element}) can perform as suitable interpolating 
functions. 

In the absence of external disturbances and assuming 
$\zvec^n_{\kvec }=\exp(-{\rm i}\omega t){\bf \zeta}^n_{\kvec }$, 
equation (\ref{eq:linearized-CBE-nth-element}) gives the exact relation 
\begin{eqnarray}
\Evec_{\kvec }(n,\Jvec)
\cdot {\bf \zeta}^{n}_{\kvec }=
\left ( \frac{ \kvec \cdot 
\frac{\partial f_0}{\partial \Jvec}}
{\kvec \cdot {\bf \Omega}-\omega } \right )
\!\! \sum_{m=-\infty}^{+\infty} \!\!\!
\Psivec_{-\kvec }(-m,n,\Jvec)
\cdot \avec^{n}_{m}.
\label{eq:formula-for-E}
\end{eqnarray}
Denoting $\langle \avec | \bvec \rangle$ 
as the inner product of the vectors $\avec$ and 
$\bvec$, one can derive the following identity 
\begin{eqnarray}
\Evec_{\kvec }(n,\Jvec)
\langle {\bf \zeta}^{n}_{\kvec } | {\bf \zeta}^{n}_{\kvec }
\rangle  \!\!\! &=& \!\!\! \left ( \kvec \cdot 
\frac{\partial f_0}{\partial \Jvec} \right ) 
\left ( \kvec \cdot {\bf \Omega}-\omega  \right )^{-1}  \nonumber \\
\!\!\! &\times& \!\!\! \sum_{m=-\infty}^{+\infty} 
\Psivec_{-\kvec }(-m,n,\Jvec)
\langle \avec^{n}_{m} | {\bf \zeta}^{n}_{\kvec } \rangle.
\label{eq:formula-for-E-right-multiplied}
\end{eqnarray}
Without loss of generality, I choose ${\bf \zeta}^{n}_{\kvec }$ 
so that $\langle {\bf \zeta}^{n}_{\kvec } | 
{\bf \zeta}^{n}_{\kvec } \rangle$ is normalised 
to unity, and take the $\omega \rightarrow 0$ limit of 
(\ref{eq:formula-for-E-right-multiplied}) to obtain a class of 
interpolating vectors $\Evec_{\kvec }(n,\Jvec)$.

To determine the constant coefficients $\langle \avec^{n}_{m} 
| {\bf \zeta}^{n}_{\kvec } \rangle$, I set $\omega=0$ and
substitute from (\ref{eq:formula-for-E}) into (\ref{eq:fundamental-DF-dens}). 
The weighted residual form of the resulting equation becomes:
\begin{eqnarray}
\!\!\!\! &{}& \!\!\!\! \textsf{\textbf{K}}(n') \cdot 
\bvec^{n'}_{m'} = (2\pi)^2 \sum_{m=-\infty}^{+\infty} 
\sum_{n=1}^{N} \Big [ \sum_{\kvec} \nonumber \\
\!\!\!\! &\times& \!\!\!\! \int  
\frac{ \kvec \cdot 
\frac{\partial f_0}{\partial \Jvec}}
{\kvec \cdot {\bf \Omega} } ~
\Psivec^{\rm T}_{\kvec }(m',n',\Jvec)
\cdot \Psivec_{-\kvec }(-m,n,\Jvec) ~ {\rm d}^2\Jvec
\Big  ]
\cdot \avec^{n}_{m},
\label{eq:bnm-anm-relation}
\end{eqnarray}
for $n'$=$1,2,\cdots,N$ and $m'\in(-\infty,+\infty)$. Assembling 
this system of matricial equations and combining it with 
(\ref{eq:reduced-poisson-all}) yield
\begin{eqnarray}
\textit{\textbf{p}}_{m'} \!\!\! &=& \!\!\!
\sum_{m=-\infty}^{+\infty} \tilde{ \textsf{\textbf{S}} }(m',m) \cdot \textit{\textbf{p}}_m   
\Longleftrightarrow \textsf{\textbf{I}} \cdot \textit{\textbf{p}}=\textsf{\textbf{S}} 
                    \cdot \textit{\textbf{p}}, \label{eq:find-a-SVD} \\
\textit{\textbf{p}} \!\!\! &=& \!\!\!  \left [ 
\begin{array}{lllllll} 
\cdots & \textit{\textbf{p}}^{\rm T}_{-2} & \textit{\textbf{p}}^{\rm T}_{-1} & 
\textit{\textbf{p}}^{\rm T}_0 & \textit{\textbf{p}}^{\rm T}_{+1} & 
\textit{\textbf{p}}^{\rm T}_{+2} & \cdots
\end{array}
\right ]^{\rm T},     \nonumber
\end{eqnarray}
where $\textsf{\textbf{I}}$ is the identity matrix and the constant 
$N_{\rm t}\times N_{\rm t}$ matrices $\tilde{\textsf{\textbf{S}}}(m',m)$ 
constitute the blocks of $\textsf{\textbf{S}}$. A non-trivial solution of 
(\ref{eq:find-a-SVD}) for $\textit{\textbf{p}}$, which will lie in the 
nullspace of $\textsf{\textbf{I}}-\textsf{\textbf{S}}$, can be obtained 
by singular value decomposition of $\textsf{\textbf{I}}-\textsf{\textbf{S}}$ 
(see Press et al. 2001). Consequently, one can compute 
$\langle \avec^{n}_{m} | {\bf \zeta}^{n}_{\kvec }\rangle$ 
and fully determine the interpolating vector. I remark that some resonant orbits 
can contribute a singularity to 
$\Evec_{\kvec }(n,\Jvec)$ 
when $\kvec \cdot {\bf \Omega}\approx 0$, but such a 
singularity is integrable. Initially axisymmetric discs satisfy the 
condition (\ref{eq:psi-for-axisymmetric}) and the interpolating 
vector reads
\begin{eqnarray}
\Evec_{\kvec }(n,\Jvec)
= \left ( \kvec \cdot 
\frac{\partial f_0}{\partial \Jvec} \right)
\left ( \kvec \cdot {\bf \Omega}  \right )^{-1}
\Psivec_{-\kvec }(-m,n,\Jvec),
\label{eq:formula-for-E-axisymmetric}
\end{eqnarray}
which has been used throughout this paper. Using 
(\ref{eq:interpolating-fncs-fourier-expansion}) one can 
verify that the elements of $\Psivec_{\kvec }(m,n,\Jvec)$ satisfy 
the continuity conditions (\ref{eq:equate-nodal-Ek}) and 
(\ref{eq:annul-nodal-Ek}), so do the interpolating functions. 

\begin{table*} 	
\caption{Eigenfrequencies of the cutout Mestel discs for $m=2$.}
\label{table1}

\begin{tabular}{ccccccc}

\hline
   &   &   &   &  &  Finite Element  &  Zang-Toomre \\   
\hline   
mode & $M_{\rm in}$ & $N$ &  $N_{\rm d}$ & Discretisation Rule 
             &  $(\Omega_{\rm p},s)$  &   $(\Omega_{\rm p},s)$ \\
\hline 
A  &  4  &  10   & 2 & $R_n=-3\ln u_n$ &  (0.450,0.189) & (0.439,0.127)   \\
A  &  4  &  20   & 2 & $R_n=-2\ln u_n$ &  (0.444,0.137) & (0.439,0.127)   \\
A  &  4  &  75   & 2 & $R_n=-1.5\ln u_n$ &  (0.442,0.127) & (0.439,0.127)   \\ \\

A  &  16  &  75  & 2 & $R_n=-1.5\ln u_n$ &  (0.486,0.316) & (0.482489,0.321296)   \\
B  &  16  &  75  & 2 & $R_n=-1.5\ln u_n$ &  (0.395,0.168) & (0.386203,0.171397)   \\
C  &  16  &  75  & 2 & $R_n=-1.5\ln u_n$ &  (1.180,0.162) & (1.161724,0.154866)   \\ \\

A  &  4  &  50   & 3 & $R_n=2(1-u_n)/u_n$ &  (0.43960,0.12675) & (0.439426,0.127181)   \\ 
 
\hline

\end{tabular}

\end{table*}

\section{Solved Examples}

To show the power of the FEM, I solve two problems regarding the dynamics 
of disc galaxies and spiral structure formation. As my first case study, 
I investigate the stability of the cutout stellar Mestel disc \citep{Z76,T77,ER98a,ER98b}. 
I have chosen this model because its shallow density falloff (like $R^{-1}$) 
fuels growing perturbations over a large radial distance from the galactic 
centre, and the methods that rely on basis function expansions 
\citep{K77,JH05,J07} require too many terms to guarantee the convergence 
of the associated eigenvalue problem. The second problem studied here is 
the disturbances induced by a satellite galaxy on its initially axisymmetric 
primary. The satellite galaxy is assumed to live inside the same dark matter 
halo of the primary, and it moves on a rosette orbit, coplanar with 
the primary's disc. 

\subsection{Stability of the stellar Mestel disc}

The equilibrium surface density and its associated 
self-gravitational potential of the Mestel disk are 
given, respectively, by \citep{ER98a}
\begin{eqnarray}
\Sigma_0(R) \!\! &=& \!\! \Sigma_s \left (\frac{R_0}{R} \right ), \\
\Phi_0(R) \!\! &=& \!\! v_0^2 \ln \left ( \frac{R}{R_0} \right ),~~
v_0^2=2\pi G\Sigma_s.
\end{eqnarray}
Here, $\Sigma_s$ is a normalising factor, $R_0$ is a length scale, 
and $v_0$ is the (constant) velocity of stars on circular orbits. 
The space of the orbital frequencies $(\Omega_1,\Omega_2)$ of the 
Mestel disc is an angular sector that extends to infinity due to 
the singularity of the gravitational force as $R\rightarrow 0$. 
The lower and upper boundaries of the frequency space are the straight 
lines $\Omega_2=\Omega_1/2$ and $\Omega_2=\Omega_1/\sqrt{2}$ that 
correspond to radial and circular orbits, respectively.

I follow \cite{Z76} and \cite{ER98a}, and use the equilibrium DF:
\begin{eqnarray}
f_0(E,L)=\frac{\Sigma_s (\gamma+1)^{1+\gamma/2}}
         {2^{\gamma/2}\sqrt{\pi}R_0^{\gamma}v_0^{\gamma+2}
	  \Gamma\left [\frac{1}{2}(\gamma+1) \right ]} 
	  L^{\gamma} e^{-(\gamma +1)E/v_0^2},
	  \label{eq:DF-Zang}
\end{eqnarray}
and the cutout DFs derived from that as 
\begin{eqnarray}
f_{\rm cut}(E,L)=
f_0(E,L)\frac{ L^{M_{\rm in}} }
{\left [L^{M_{\rm in}}+(R_0 v_0)^{M_{\rm in}} \right ] 
 },
 \label{eq:inner-cutout-DF}
\end{eqnarray}
where the integer exponent $M_{\rm in}$ specifies the sharpness 
of the inner cutout. The cutout DF introduced in (\ref{eq:inner-cutout-DF}) 
means that stars with $L \ll R_0v_0$ will not participate in density 
perturbations though they can still contribute to the mean-field 
gravitational potential.

\begin{figure*}
\centerline{\hbox{\includegraphics[width=0.31\textwidth]{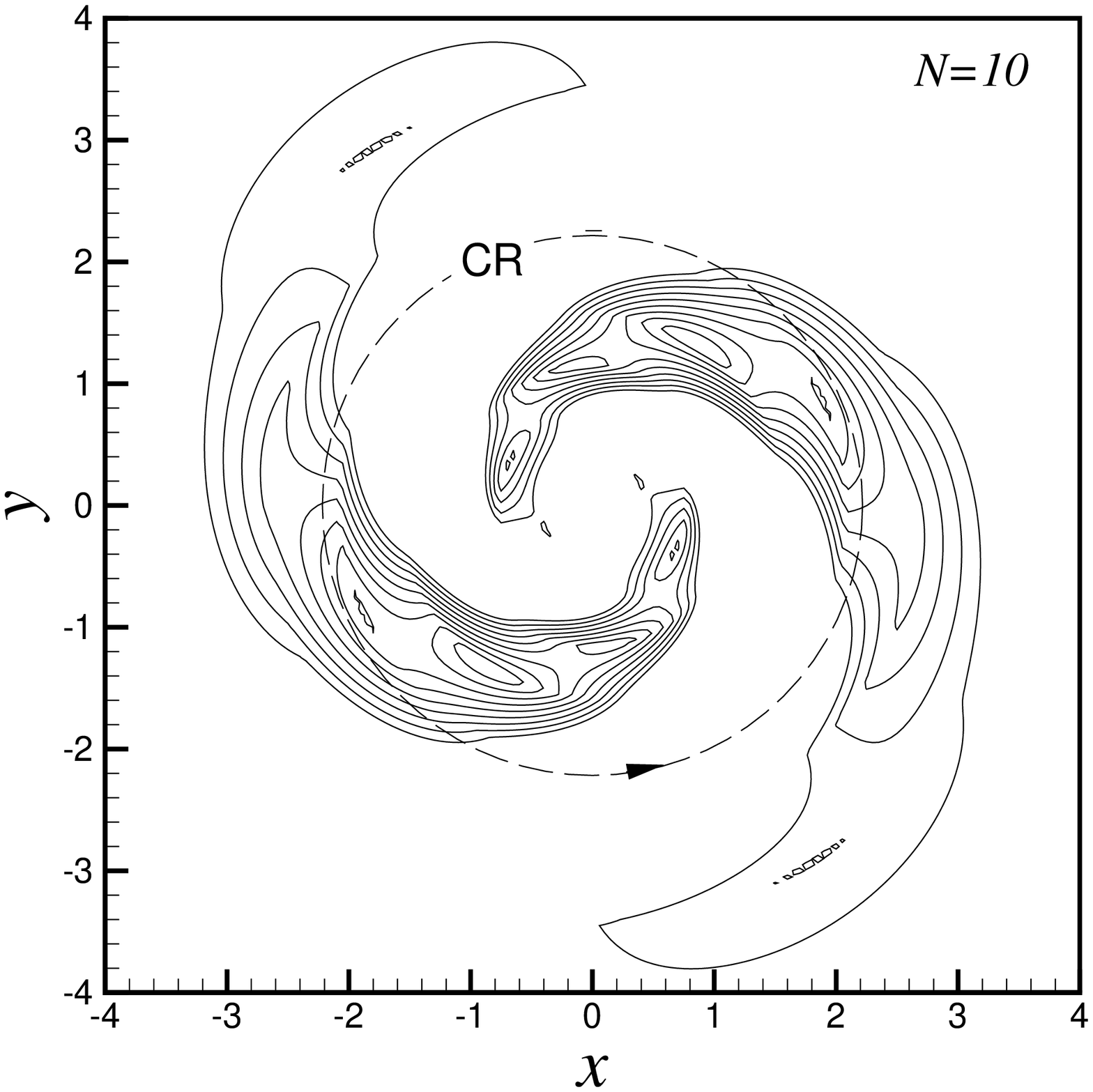}}
            \hbox{\includegraphics[width=0.31\textwidth]{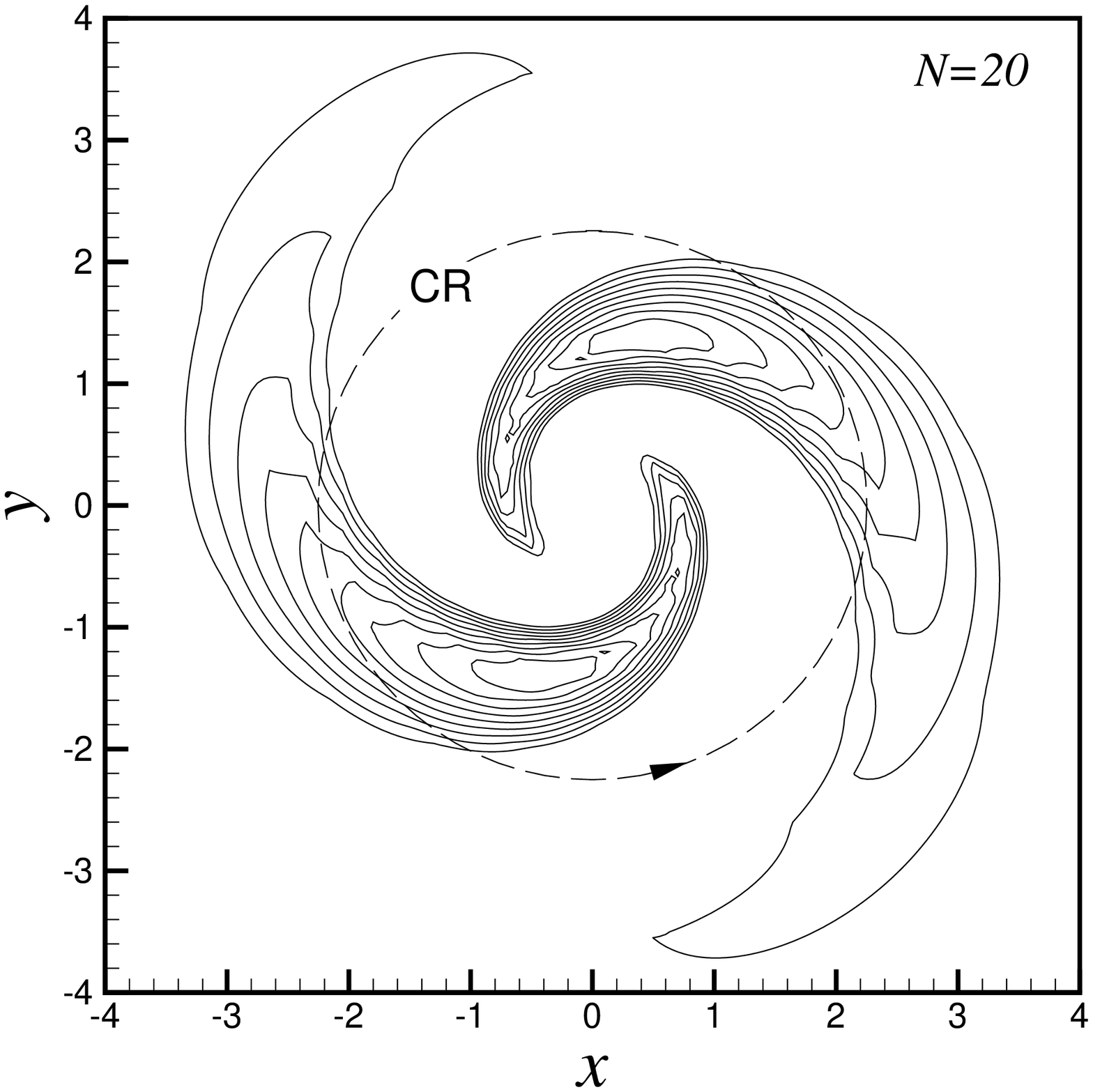}} 
            \hbox{\includegraphics[width=0.31\textwidth]{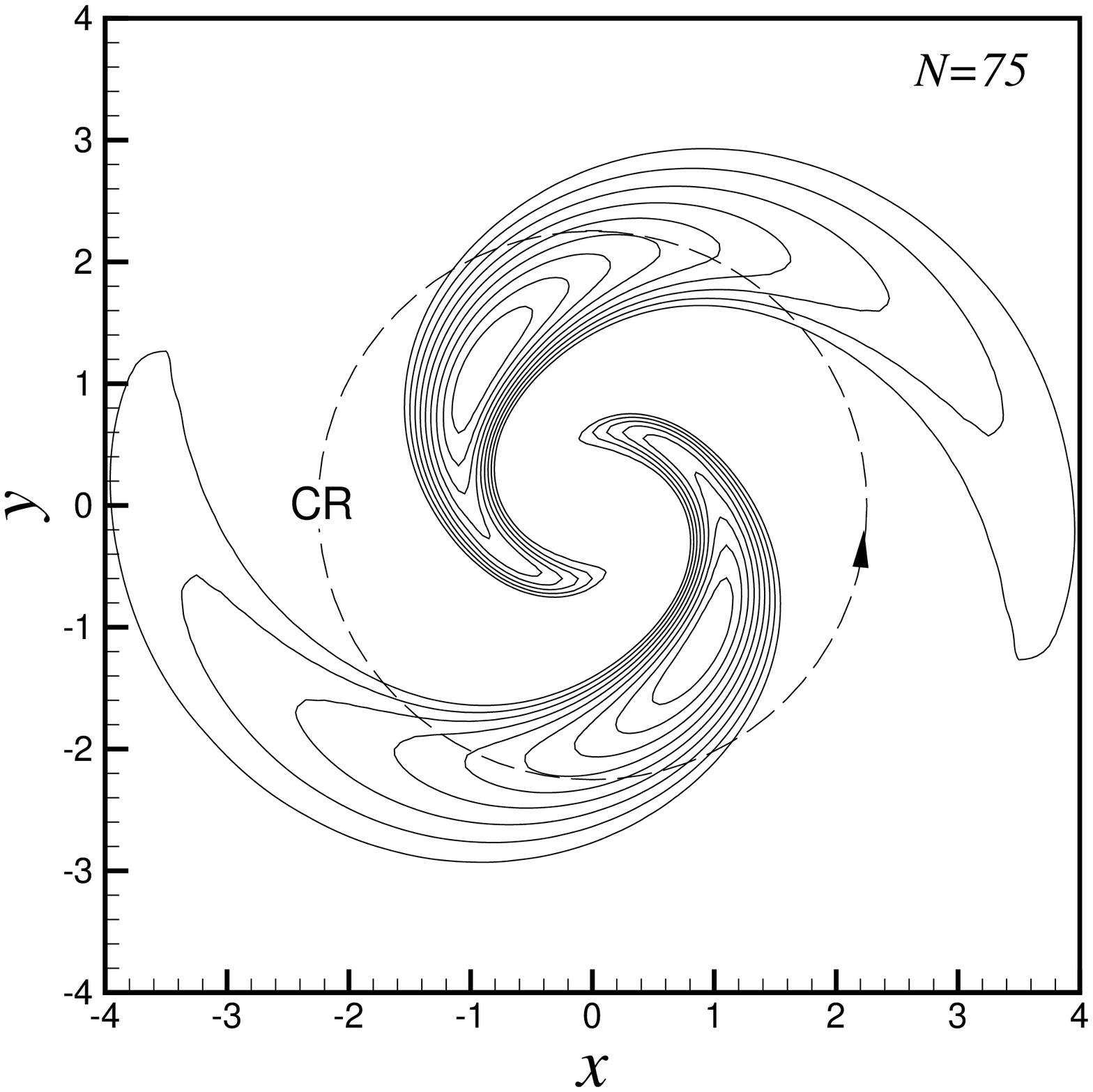}}
	    }
\centerline{\hbox{\includegraphics[width=0.31\textwidth]{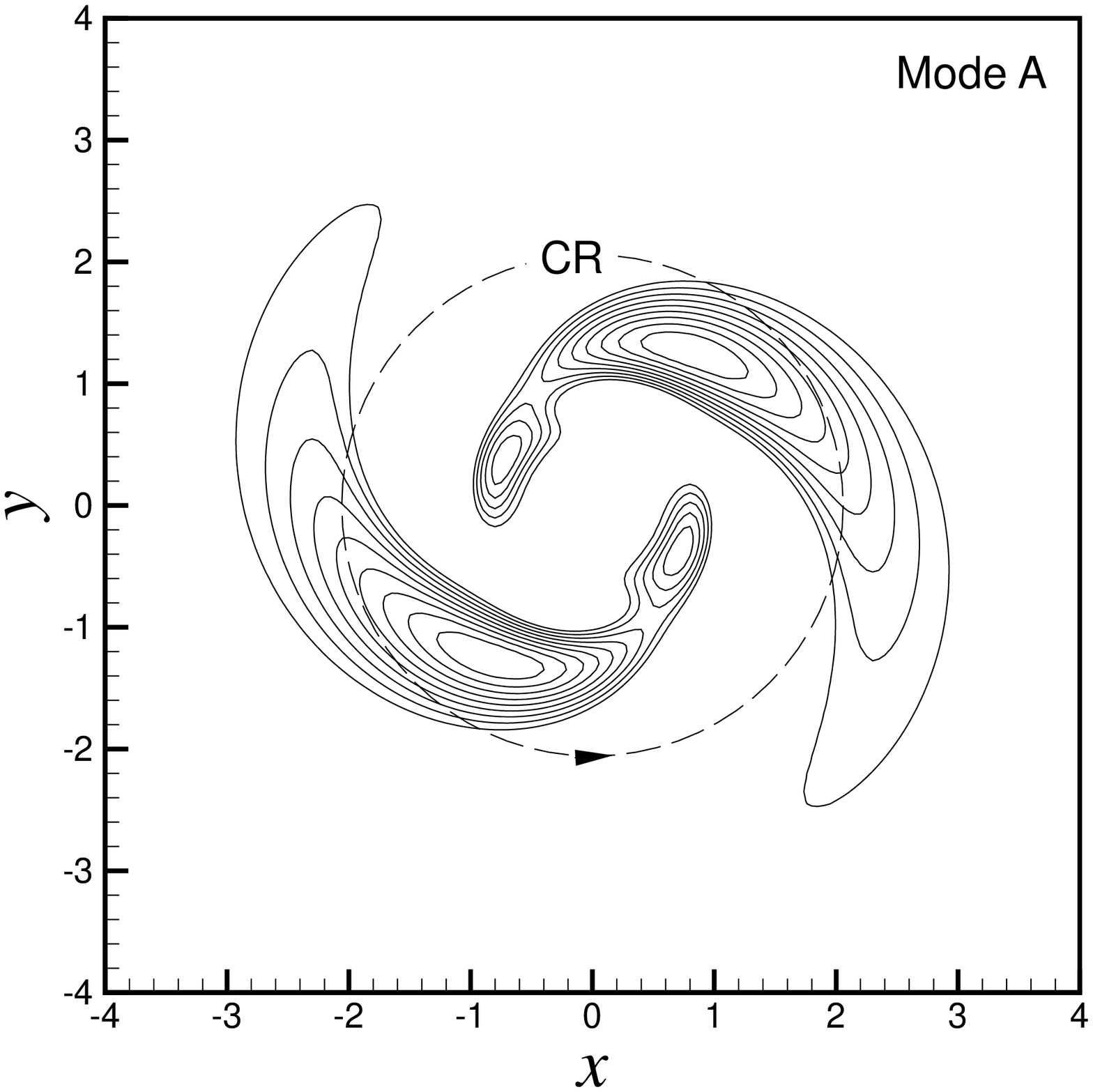}} 
            \hbox{\includegraphics[width=0.31\textwidth]{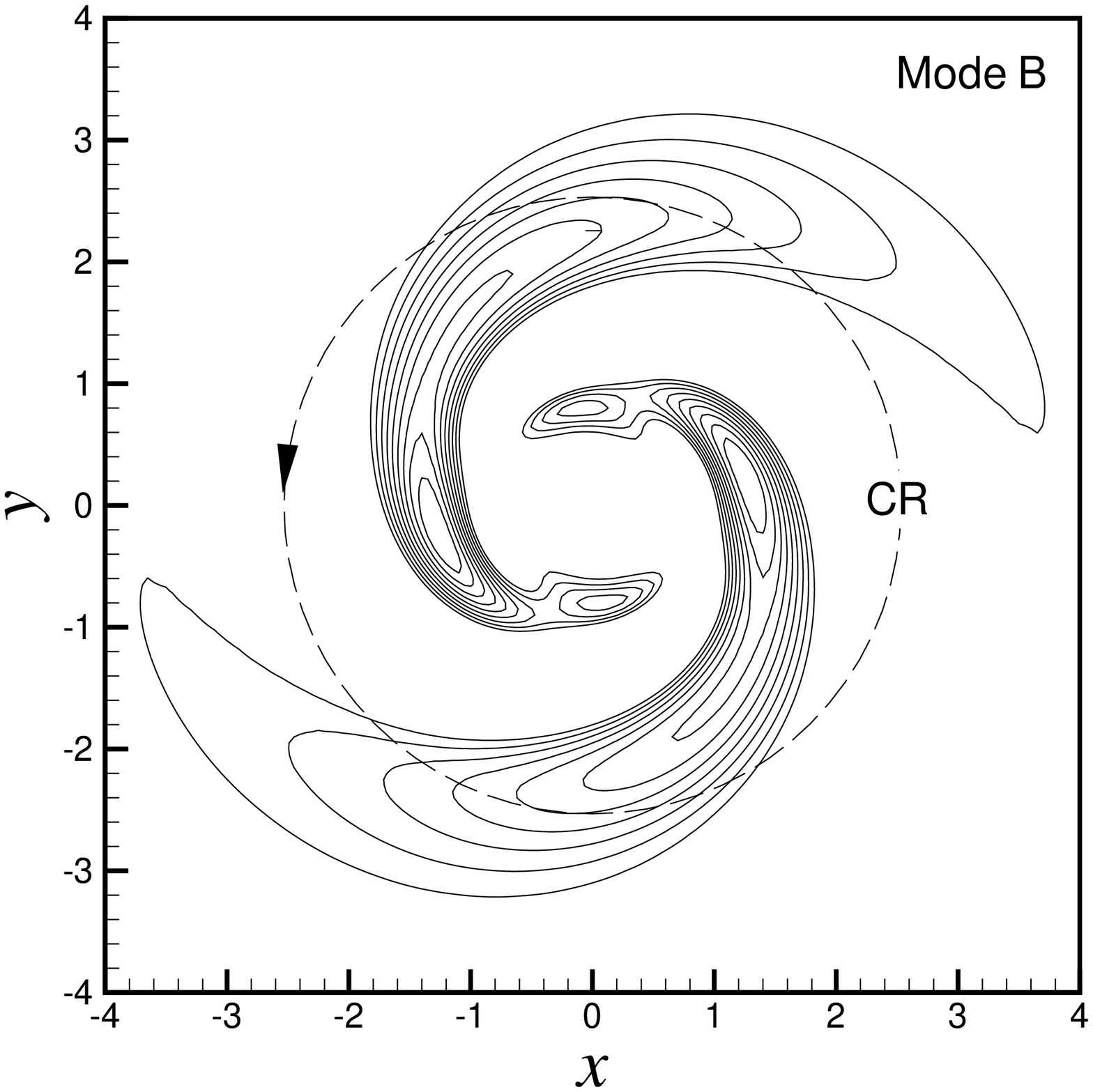}}
            \hbox{\includegraphics[width=0.31\textwidth]{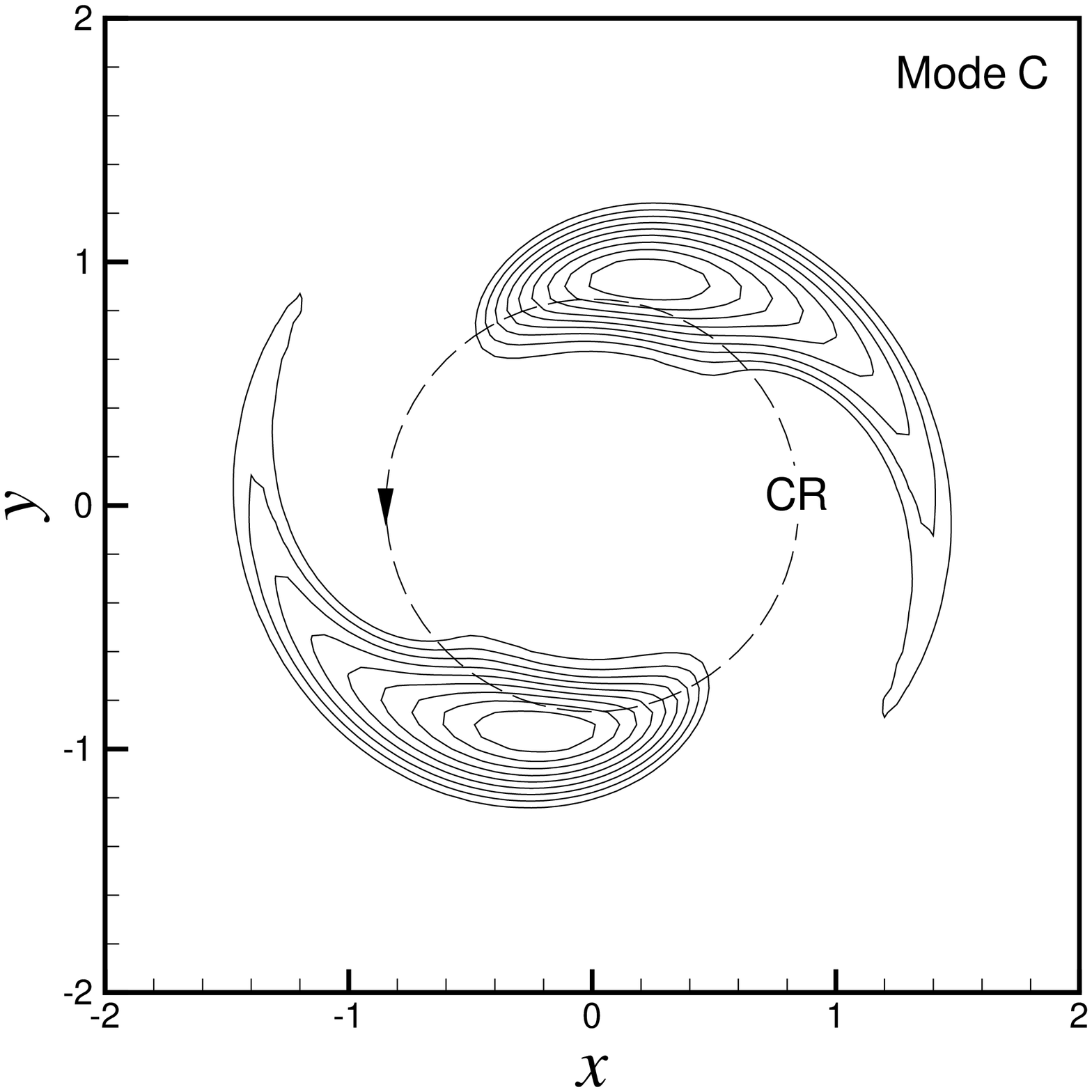}}	    
	    }
\caption{{\em Top row:} The fundamental unstable mode A of the cutout Mestel disk 
with $M_{\rm in}=4$, $N_{\rm d}=2$ and $\gamma=6$. Left, middle and right figures 
correspond to $(N,\alpha_1)=(10,3)$, $(N,\alpha_1)=(20,2)$ and $(N,\alpha_1)=(75,1.5)$, 
respectively. It is evident that increasing the number of ring elements smoothens 
the density isocontours. This mode has also been displayed in Figure 12 of \citet{T77}. 
{\em Bottom row:} Modes A, B and C of the cutout Mestel disc with 
$(M_{\rm in},\alpha_1)=(16,1.5)$, $N=75$ and $\gamma=6$. Mode C is an inner edge 
mode that develops where the cutout surface density has a rising profile versus 
$R$. In all figures, positive isodensity contours of $\Sigma_1(R,\phi,0)$ have 
been plotted from 10$\%$ to 90$\%$ of the maximum, with the steps of 10$\%$. 
Dashed circles mark the location of the corotation resonance (CR).}
\label{fig2}
\end{figure*}

I set $R_0$=$v_0$=$G$=1 and choose a model with 
$(M_{\rm in},\gamma)=(4,6)$. Since most perturbations decay 
monotonically as $R\rightarrow \infty$, it is useful to adopt 
a nonuniform mesh so that the widths of elements decrease towards 
the disc centre. I divide the configuration space to $N$ finite 
ring elements and apply either of the following rules 
\begin{eqnarray}
R_n=-\alpha_1 \ln u_n,~~R_n = \alpha_2 \frac{1-u_n}{u_n},
\end{eqnarray}
to generate elements of nonuniform width where 
\begin{eqnarray}
u_n=1-\frac{1}{2(N+1)}-\frac{n-1}{N+1},~~n=1,2,\cdots,N+1.
\end{eqnarray}
All integrals over the action space are evaluated 
after carrying out three successive changes of 
variables as
\begin{eqnarray}
(J_1,J_2) \rightarrow (E,L) \rightarrow
(R_{\rm min},R_{\rm max}) \rightarrow (R_c,e).
\end{eqnarray}
Here $R_{\rm min}$ and $R_{\rm max}$ are the minimum and 
maximum galactocentric distances of rosette orbits and
\begin{eqnarray}
R_{\rm min} \!\! &=& \!\! R_c(1-e),~~R_{\rm max}=R_c(1+e),
~~ 0\le e \le 1.
\end{eqnarray}
For axisymmetric discs, the Fourier numbers $\kvec=(k_1,k_2)$ 
contract to $k_1=0,\pm 1,\pm 2,\cdots$ and $k_2=m$, which 
substantially decreases the size of the eigensystem 
(\ref{eq:linear-eigensystem}). Nevertheless, one must 
truncate the Fourier series in terms of the radial angle 
$w_1$. Taking $-10 \le k_1 \le 10$ suffices in most systems 
(e.g., Jalali \& Hunter 2005). The eigenfrequencies and 
their corresponding eigenvectors of equation 
(\ref{eq:linear-eigensystem}) are computed using the 
same algorithms and subroutines of \citet{J07}.  

Top row in Figure \ref{fig2} shows the fastest growing 
bisymmetric mode A of $m=2$, which has been obtained using 
the Galerkin form (\ref{eq:perturbed-CBE-projected-nth-element})
and through solving (\ref{eq:linear-eigensystem}) for different 
finite element gridings by taking $-2\le k_1\le 5$. 
The pattern speed $\Omega_{\rm p}$ and growth rate $s$ of 
this mode (note: $\omega=m\Omega_{\rm p}+{\rm i}s$) are given 
in Table \ref{table1} up to three decimal places, and they 
are compared with the values computed by \cite{Z76} and also 
reported in \cite{T77}. It is evident that the FEM with 
$(N,\alpha_1)=(75,1.5)$ has given a satisfactory accuracy 
of $0.6\%$. By increasing the number of ring elements the 
mode shape is smoothened too. I have also utilised 
(\ref{eq:perturbed-CBE-projected}) and computed the eigenfrequency 
spectrum of a model with $(M_{\rm in},\gamma)=(16,6)$. 
The spectrum includes three prominent modes A, B and C, 
which have been found with an average error of $2\%$ 
(see Table \ref{table1}). Mode C is an inner edge mode that 
comes into existence due to sharp cutout. The eigenfrequencies
of all these modes have been known to Alar Toomre up to six
decimal places (private communication; see Table \ref{table1}) 
and I have demonstrated their corresponding mode shapes in 
the bottom row of Figure \ref{fig2}. They have not already 
been displayed in the literature. 

The accuracy of FEM calculations is determined by several 
factors: (i) element types, interpolation rule and the degree
of differentiability at the nodes (ii) the number of Fourier 
terms in the $w_1$-direction (iii) the number of ring elements 
and their sizes specified by $\Delta R_n$ (iv) the precision 
of the integrals taken over the action space in constructing 
the matrices $\textsf{\textbf{L}}$, $\textsf{\textbf{E}}_1$, 
$\textsf{\textbf{E}}_2$ and $\textsf{\textbf{E}}_3$ (v) the 
accuracy of eigenvalue solver. Furthermore, not all eigenmodes 
will have the same precision because their clumps do not 
simultaneously fall in a region with appropriate number of 
elements. For instance, the growth rate of mode C is less 
accurate than modes A and B, and the pattern speeds of all 
modes have been overestimated. These errors correlate mainly 
with the type and sizes of elements. Using the discretisation 
rule $R_n=2(1-u_n)/u_n$, which increases the radius of the 
outermost element, and by applying quadratic elements 
($N_{\rm d}=3$), the accuracy of $(\Omega_{\rm p},s)$ for mode A 
of the model with $M_{\rm in}=4$ reaches to an impressive 
level of $(\pm 0.0002,\pm 0.0004)$ even by taking $N=50$ elements. 
Last row in Table \ref{table1} compares the eigenfrequency found 
by FEM with Alar Toomre's recent high-precision results. 
Although an adaptive mesh refinement can enhance the accuracy of 
an individual mode (if not the whole spectrum), a significant 
improvement is anticipated only by a $C_1$ finite element 
formulation that assures the smoothness of perturbed quantities
over the entire configuration and phase spaces. 

\begin{figure*}
\centerline{\hbox{\includegraphics[width=0.32\textwidth]{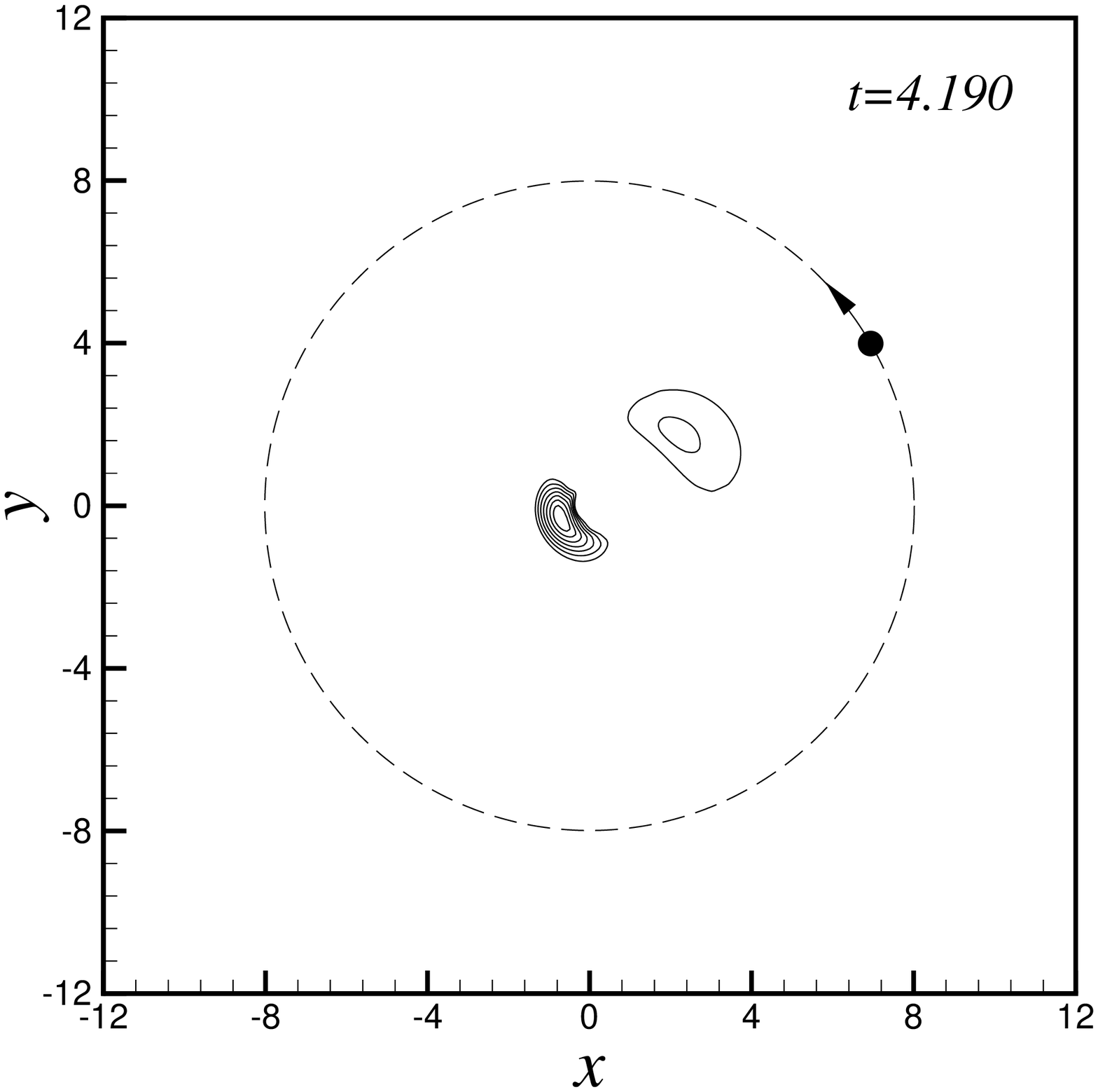}} 
            \hbox{\includegraphics[width=0.32\textwidth]{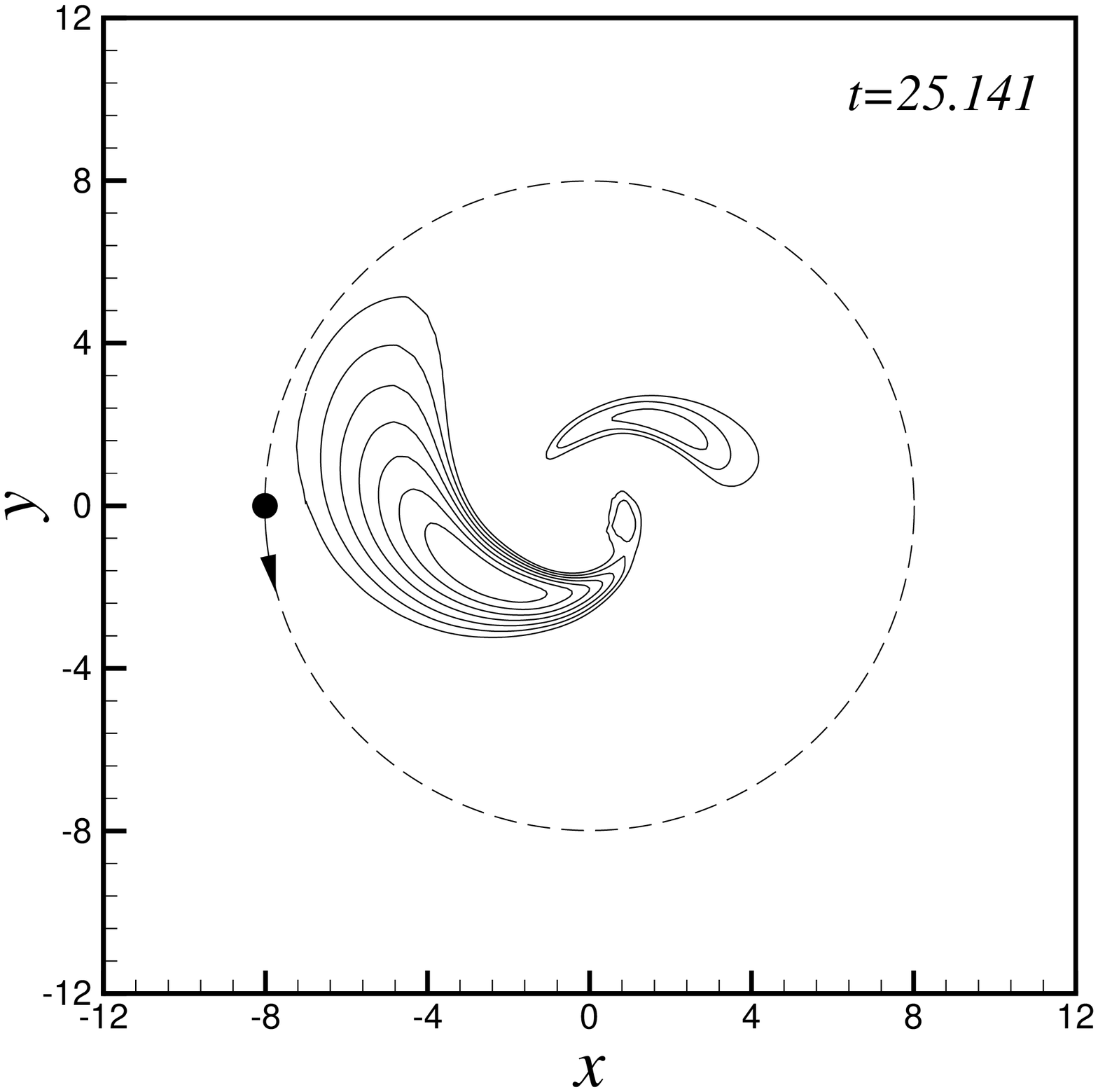}}
            \hbox{\includegraphics[width=0.32\textwidth]{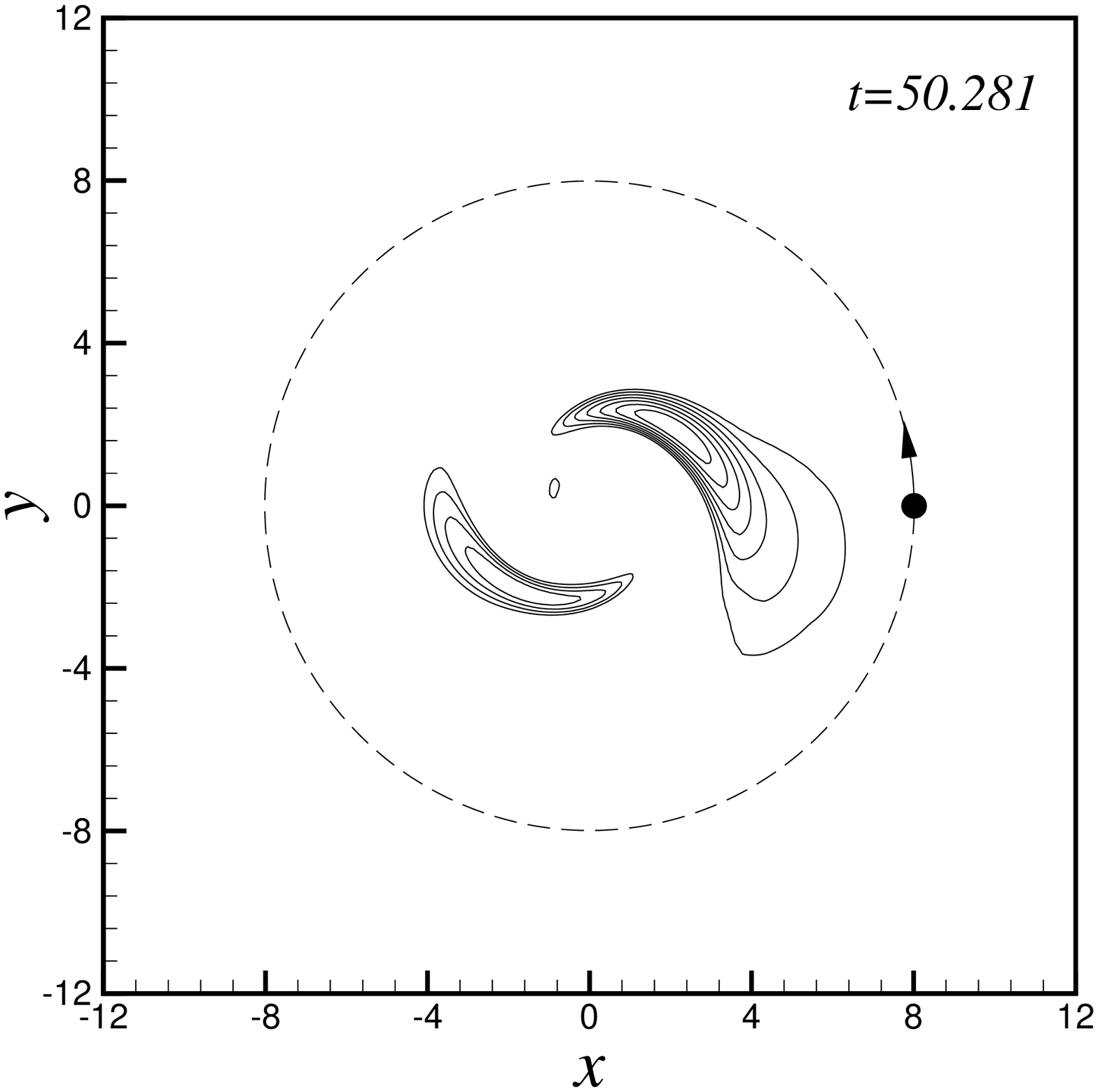}}	    
            }
\centerline{\hbox{\includegraphics[width=0.32\textwidth]{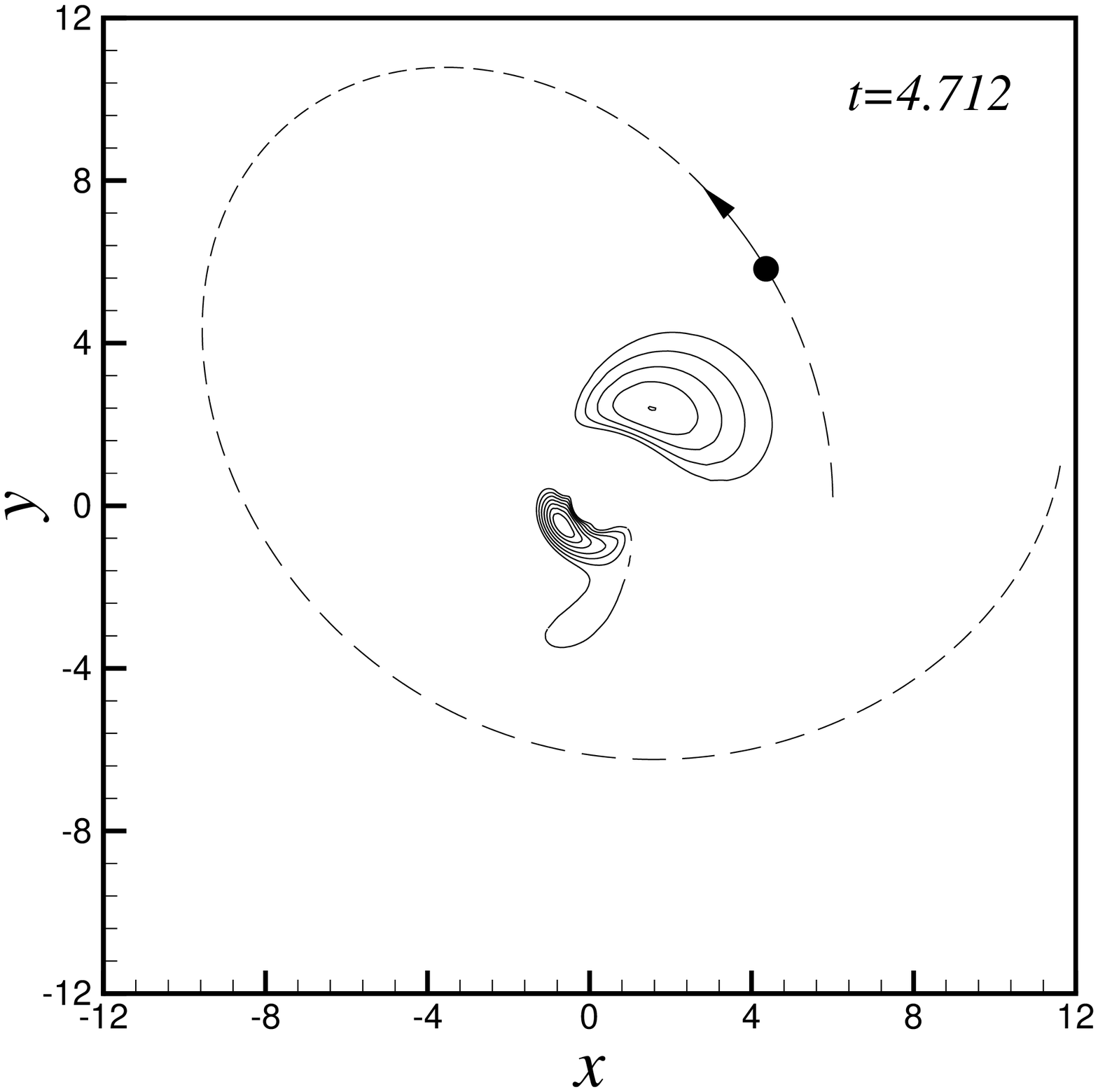}} 
            \hbox{\includegraphics[width=0.32\textwidth]{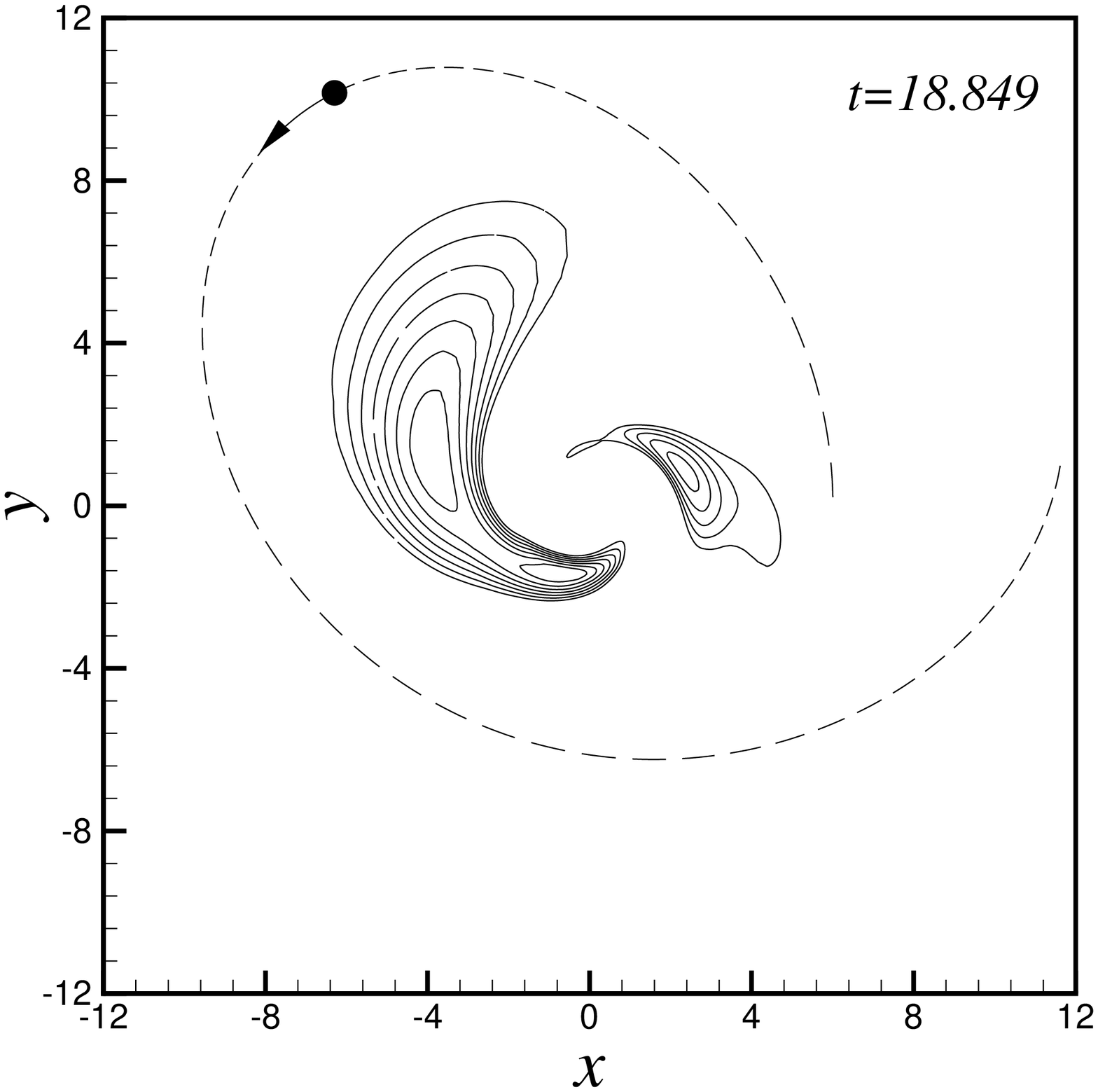}}
            \hbox{\includegraphics[width=0.32\textwidth]{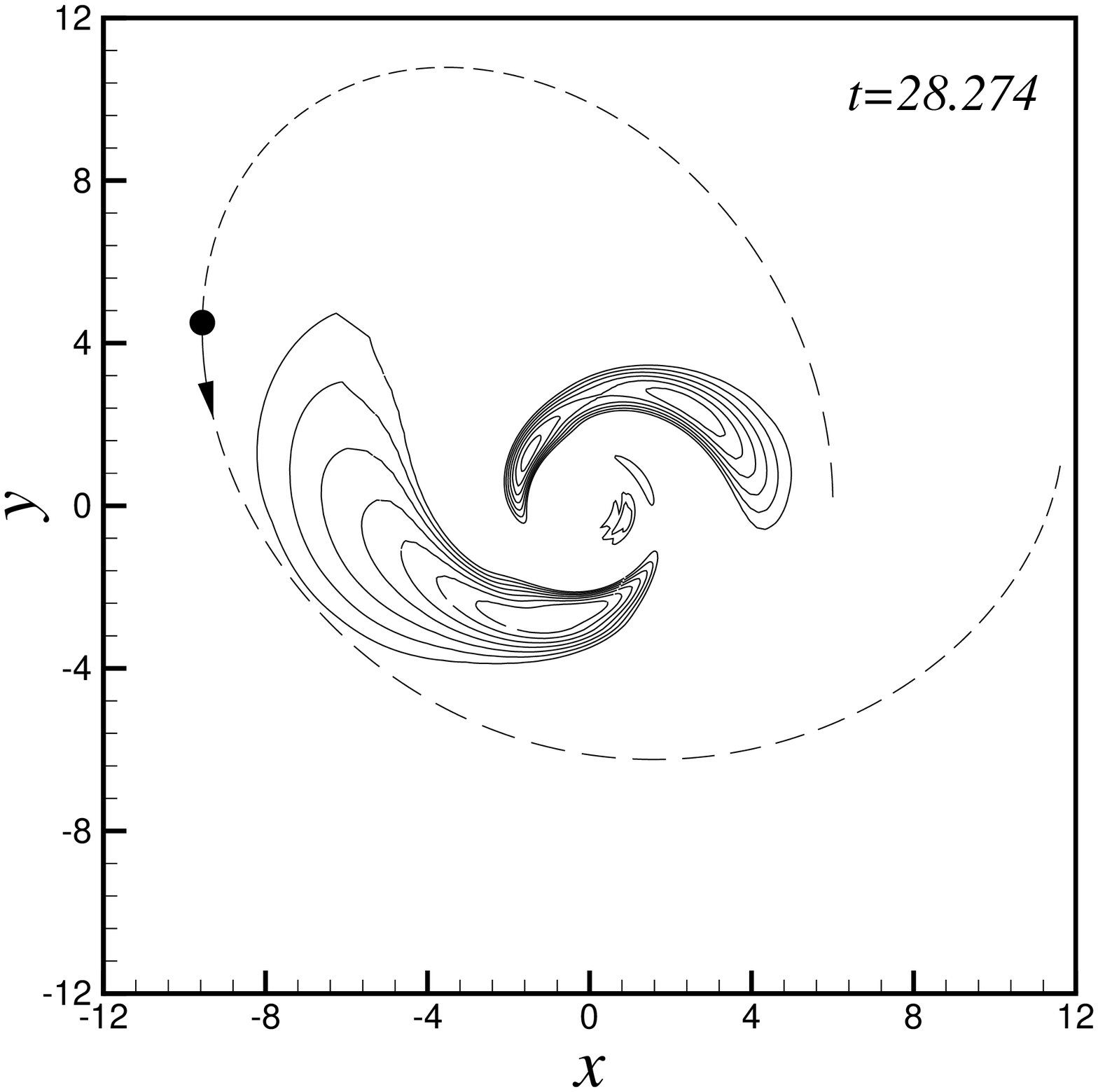}}	    
            }
\caption{Disturbances induced by a satellite galaxy (filled circle) 
on a stable doubly cutout Mestel disc. {\em Top row}: The satellite moves on 
a circular orbit of radius $R_{\rm S}=8$. {\em Bottom row}: Satellite's orbit 
is a rosette of $R_{\rm min}=6$ and $R_{\rm max}=12$. The initial azimuth 
$\phi_{\rm S}(0)$ of the perturber is zero. This state of the system also 
defines the origin of time: $t=0$. Only positive isodensity contours 
of $\Sigma_1(R,\phi,t)$ have been plotted from 30$\%$ to 90$\%$ of the 
maximum, with the steps of 10$\%$.}
\label{fig3}
\end{figure*}

\subsection{Perturbations induced by a satellite galaxy}

Kinematics and dynamics of galaxies are highly influenced by their
environment. Mergers, close encounters, and bound companions determine 
the structure and evolution of most cluster galaxies. The FEM developed 
in this study is capable of modelling the disturbances of complex 
interactions between stellar systems, and it can complement $N$-body 
simulations in the modeling of multi-scale structure formation and 
evolution of galaxies. As an illustrative example, I apply the FEM 
and investigate the induced disturbances of a stellar disc by a distant 
satellite galaxy. The primary stellar disc is assumed to be a doubly 
cutout Mestel disc with the DF:
\begin{eqnarray}
f_{\rm cut}(E,L)=
\frac{ \beta f_0(E,L) L^{M_{\rm in}} L_c^{M_{\rm out}}}
{\left [L^{M_{\rm in}}+(R_0 v_0)^{M_{\rm in}} \right ] 
 \left [L^{M_{\rm out}}+L_c^{M_{\rm out}} \right ]
 },
 \label{eq:doubly-cutout-DF}
\end{eqnarray}
where $0 < \beta \le 1$, and $f_0$ has been defined in (\ref{eq:DF-Zang}). 
Immobilised particles with $L \ll R_0v_0$ and $L\gg L_c$ simulate, respectively, 
a hot bulge and a rigid dark halo. Unstable modes are indeed the homogeneous 
solutions of equation (\ref{eq:perturbed-CBE-projected-assembled}). 
By adjusting $\beta$ and the set of parameters $(L_c,M_{\rm in},M_{\rm out})$, 
one can build a stable axisymmetric disc and study only the effect of 
the external perturber as the particular solutions of 
(\ref{eq:perturbed-CBE-projected-assembled}).

I make two simplifying assumptions for the motion of the 
satellite galaxy: (i) the dynamical friction of both the 
baryonic and dark matter components is ignored (ii) the 
satellite galaxy is a point mass that keeps moving on a 
rosette orbit while its motion is governed by the same 
massive dark halo that hosts the primary. One can therefore 
neglect the indirect gravitational force of the satellite 
galaxy on the disc stars and write the disturbance 
function as
\begin{eqnarray}
\Phi_{\rm e}(R,\phi,t)=-\frac{GM_{\rm S}}{R_{\rm S}} 
\sum_{i=0}^{\infty} \left ( \frac{R}{R_{\rm S}} \right )^i P^0_i
\left [\cos (\phi-\phi_{\rm S}) \right ],
\label{eq:disturbance-function-expansion}
\end{eqnarray}
where $[R_{\rm S}(t),\phi_{\rm S}(t)]$ are the polar coordinates 
of a satellite (of mass $M_{\rm S}$) measured with respect to a 
non-rotating frame whose origin is attached to the primary's 
centre. The potential field of the rigid dark halo is logarithmic
at distant regions. 
Thus, the motion of the satellite is governed by
\begin{eqnarray}
\frac{{\rm d}^2 R_{\rm S}}{{\rm d}t^2}-R_{\rm S}
\left ( \frac{{\rm d}\phi_{\rm S}}{{\rm d}t} \right )^2=-\frac{v_0^2}{R_{\rm S}},
~~~
\frac{{\rm d}}{{\rm d}t}\left [
R^2_{\rm S} \left ( \frac{{\rm d}\phi_{\rm S}}{{\rm d}t} \right ) \right ]=0.
\end{eqnarray}
To implement the FEM, one needs to represent the disturbance function 
in terms of the angle-action variables. I define 
\begin{eqnarray}
X_{1,k_1}(\Jvec) \!\!\! &=& \!\!\!
\frac{1}{2\pi} \oint R \cos [k_1 w_1+(w_2-\phi)] ~ {\rm d}w_1, \\
Y_{j,k_1}(\Jvec) \!\!\! &=& \!\!\!
\frac{1}{2\pi} \oint R^2 \cos [k_1 w_1+j(w_2-\phi)] ~ {\rm d}w_1,
\end{eqnarray}
and keep the leading $i\le 2$ terms of (\ref{eq:disturbance-function-expansion}) 
to obtain 
\begin{eqnarray}
\Phi_{\rm e}(\wvec,\Jvec,t) \!\!\!\! &\approx & \!\!\!\!
{\rm Re} \Bigg [ -\frac{GM_{\rm S}}{R_{\rm S}}
-\frac{GM_{\rm S}}{4R_{\rm S}^3} \!\!\!  \sum_{k_1=-\infty}^{+\infty}
Y_{0,k_1}(\Jvec) e^{{\rm i}k_1 w_1} \nonumber \\
\!\!\!\! &-& \!\!\!\! \frac{GM_{\rm S}}{R_{\rm S}^2} 
\!\!\! \sum_{k_1=-\infty}^{+\infty} \!\! e^{-{\rm i}\phi_{\rm S}}
X_{1,k_1}(\Jvec) e^{ {\rm i}(k_1 w_1+w_2) } \nonumber \\
\!\!\!\! &-& \!\!\!\! \frac{3GM_{\rm S}}{4R_{\rm S}^3} 
\!\!\! \sum_{k_1=-\infty}^{+\infty} \!\! e^{-2{\rm i}\phi_{\rm S}}
Y_{2,k_1}(\Jvec) e^{ {\rm i}(k_1 w_1+2 w_2)} \Bigg ].
\label{eq:disturbance-function-action-angle}
\end{eqnarray}
The first term on the right hand side of (\ref{eq:disturbance-function-action-angle}) 
can be dropped because it does not contribute to the disturbing force. The 
second term generates an unsteady, axisymmetric, particular solution of 
(\ref{eq:perturbed-CBE-projected-assembled}). The third and 
fourth terms excite, respectively, rotating patterns of angular 
wavenumbers $m=1$ and $m=2$. Since I am considering the linearised CBE, 
all solutions will be superposed to get the imposed perturbed density.

To this end, I adopt the Galerkin form (\ref{eq:perturbed-CBE-projected-nth-element})
and set the model parameters to $v_0=R_0=G=1$, $\gamma=6$, $\beta=0.1$, $L_c=4$, 
and $M_{\rm in}=M_{\rm out}=2$, which result in a disc of active mass 
$M_{\rm disc}=0.417$ in the normalised units. My calculations using the 
eigensystem (\ref{eq:linear-eigensystem}) shows that this disc is stable to 
internal excitation of any angular wavenumber $m$. It is therefore guaranteed 
that in the presence of an external perturber, the disc will not develop 
an exponentially growing mode. I assume $GM_{\rm S}=0.04$, generate a 
non-uniform grid of $(N,\alpha_1)=(75,2)$, and keep the terms corresponding 
to the radial Fourier numbers $-5\le k_1\le +5$. Since I have taken only 
the first three terms of the disturbance function ($0\le i\le 2$), there 
will be $3\times 11$ unknown vectors of the amplitude functions $\zvec_{\kvec}(t)$, 
each being a $76\times 1$ column vector. I turn on the forcing vector 
$\textit{\textbf{Z}}(\kvec,t)$ when the satellite's true anomaly 
is $\phi_{\rm S}=0$ at the origin of time ($t=0$), and integrate equations 
(\ref{eq:perturbed-CBE-projected-assembled}) by an accuracy of $10^{-4}$ 
using the subroutine ODEINT of \citet{Press01}.
 
Top row in Figure \ref{fig3} displays three snap shots of generated spiral arms 
as the satellite moves on a circular orbit of radius $R_{\rm S}=8$. The fundamental 
feature of density perturbations is that the closer spiral arm to the perturber is 
more extensive than the arm on the opposite side. This shows the dominance of the 
$i=1$ term in (\ref{eq:disturbance-function-expansion}). Bottom row in Figure \ref{fig3} 
demonstrates the density perturbations induced by the same satellite of $GM_{\rm S}=0.04$,
but orbiting on a rosette of $R_{\rm min}=6$ and $R_{\rm max}=12$. It is evident 
that the spiral arm opposite to the satellite's location is amplified as the 
satellite descends from its orbital apocentre. For both the circular and rosette 
orbits, the major wave packets of density perturbations lead the satellite. This 
phase lead increases as the time is elapsed, but it is more prominent 
(even more than $90^\circ$) when the satellite's orbit is highly eccentric.
The results are not altered by taking $-8\le k_1\le +8$, which shows a fast 
convergence of Fourier series in terms of $w_1$. 

%\vspace{-0.4cm}

\section{Conclusions}

I modelled the dynamics of collisionless stellar systems using finite 
elements and used the FEM to study the spiral structure formation. 
The method is highly adaptable to all given initial density profiles,
and it can be applied to stability problems as well as disturbances 
induced by external sources. The FEM converges by taking a relatively 
small number of radial elements and it can accurately resolve different 
growing modes of unstable discs. Although the examples of this study 
were confined to the perturbations of axisymmetric discs, the derived 
equations are quite general and can be readily applied to elongated discs 
with rich orbital structures.

Applying a Fourier expansion in the azimuthal $\phi$-direction is favoured
in  theoretical studies of toy galaxy models. In real systems one may need 
too many Fourier terms to get converged the perturbed density and its 
corresponding potential. This problem can be avoided by using a two 
dimensional finite element grid in the ($R,\phi$)-space, instead of ring 
elements only in the radial direction. Two dimensional elements can be 
triangular, rectangular or mapped ones depending on the shape of the galaxy. 
Moreover, a two dimensional grid makes the governing equations independent 
of the wavenumber $m$, and simplifies the form of projected equations.

For models with $N<100$ ring elements, and for radial Fourier numbers in 
the range $-10 \le k_1 \le 10$, the largest size of ordinary differential 
equation that must be integrated to study the pattern evolution, is 
of ${\cal O}(10^3)$. This is far less than motion equations solved in 
$N$-body simulations of typical disc galaxies. Given the accuracy 
of results that I obtained for the global modes of the cutout Mestel disc, 
the FEM can thus be regarded as an extremely efficient technique as long as 
the integrator of evolutionary equations is concerned. However, the spectral 
analysis of orbit families and the calculation of the row vector 
$\Psivec_{-\kvec }(-m,n,\Jvec)$ is costly. 
This makes the numerical effort of the FEM comparable with Schwarzschild's 
(1979) method because one needs to identify and analyse the orbits that 
visit each element.

The error of FEM simulations can be controlled not necessarily by 
increasing the number of elements, but by suitable (adaptive) variation 
of element sizes and increasing the order of interpolating functions.
Perhaps the most remarkable advantage of the FEM is that it can be directly 
linked with computational fluid dynamics codes, which mainly use 
finite difference and finite element methods, to study the co-evolution 
of the stellar and gas components of galaxies. The elements of a compound
medium does not evolve according to the same (or similar) physical principles,
but a common simulation method can make a fruitful bridge between them. 
For instance, the FEM modeling of the stellar and gas components may 
provide a better understanding of the starburst activity in spiral arms.
The reliance of FEM on matrix algebra also distinguishes it from other 
simulation methods. Since matrix summations and products are performed 
on graphic cards more efficient than CPU, commercial hardwares in the 
PC market can be used to build special-purpose computer boards for the 
FEM simulations of complex stellar systems.

%\vspace{-0.3cm}

\section*{Acknowledgements}
I express my sincere thanks to Alar Toomre for his illuminating discussions 
and for providing me with his new high-precision results of Zang's disc. 
I also thank the anonymous referee for a useful report. This work was 
partially supported by the Research Vice-Presidency at Sharif University 
of Technology. 

%\vspace{-0.3cm}


\begin{thebibliography}{}
%    
  \bibitem[\protect\citeauthoryear{Agertz et al.}{2007}]{Agertz07}
   Agertz O., Moore B., Stadel J., Potter D. et al., MNRAS, 380, 963
%    
  \bibitem[\protect\citeauthoryear{Binney \& Tremaine}{2008}]{BT08}
    Binney J., Tremaine S., 2008, Galactic Dynamics. 2nd edition, 
    Princeton University Press, Princeton
%    
  \bibitem[\protect\citeauthoryear{Clutton-Brock}{1972}]{CB72}
    Clutton-Brock M., 1972, Ap\&SS, 16, 101
%
  \bibitem[\protect\citeauthoryear{Clutton-Brock}{1973}]{CB73}
    Clutton-Brock M., 1973, Ap\&SS, 23, 55    
%    
  \bibitem[\protect\citeauthoryear{Evans \& Read}{1998a}]{ER98a}
    Evans N.W., Read J.C.A., 1998a, MNRAS, 300, 83 
%
   \bibitem[\protect\citeauthoryear{Evans \& Read}{1998b}]{ER98b}
    Evans N.W., Read J.C.A., 1998b, MNRAS, 300, 106   
%        
  \bibitem[\protect\citeauthoryear{Gaburov, Harfst \& Portegies Zwart}{2009}]{GHPZ09}
    Gaburov E., Harfst S., Portegies Zwart S., 2009, New Astronomy, 14, 630    
%
  \bibitem[\protect\citeauthoryear{Hernquist \& Ostriker}{1992}]{HO92}
    Hernquist L., Ostriker J.P., 1992, ApJ, 386, 375
%
  \bibitem[\protect\citeauthoryear{Jalali \& Hunter}{2005}]{JH05}
    Jalali M.A., Hunter C., 2005, ApJ, 630, 804
%    
  \bibitem[\protect\citeauthoryear{Jalali}{2007}]{J07}
    Jalali M.A., 2007, ApJ, 669, 218 
%
  \bibitem[\protect\citeauthoryear{Kaasalainen \& Binney}{1994a}]{KB94a}
    Kaasalainen M., Binney J., 1994a, PRL, 73(18), 2377
%    
  \bibitem[\protect\citeauthoryear{Kaasalainen \& Binney}{1994b}]{KB94b}
    Kaasalainen M., Binney J., 1994b, MNRAS, 268, 1033     
%   
  \bibitem[\protect\citeauthoryear{Kalnajs}{1976}]{K76}
    Kalnajs A.J., 1976, ApJ, 205, 745    
%    
  \bibitem[\protect\citeauthoryear{Kalnajs}{1977}]{K77}
    Kalnajs A.J., 1977, ApJ, 212, 637
%
  \bibitem[\protect\citeauthoryear{Lewis, Nithiarasu \& Seetharamu}{2004}]{LNS04}
    Lewis R.W., Nithiarasu P., Seetharamu K.N., 2004, Fundamentals of
    the Finite Element Method for Heat and Fluid Flow, 
    John Wiley \& Sons, West Sussex, England
%
  \bibitem[\protect\citeauthoryear{Makino et al.}{2003}]{M03}
    Makino J., Fukushige T., Koga M., Namura K., 2003,
    PASJ, 55, 1163
%
  \bibitem[\protect\citeauthoryear{McGill \& Binney}{1990}]{MB90}
    McGill C., Binney J., 1990, MNRAS, 244, 634   
%
  \bibitem[\protect\citeauthoryear{Parker et al.}{2008}]{parker08}  
    Parker J., Lyzenga G., Norton C., Zuffada C., Glasscoe M., Lou J., 
    Donnellan A., 2008, Pure appl. geophys., 165, 497     
%
  \bibitem[\protect\citeauthoryear{Portegies Zwart, Belleman \& Geldof}{2007}]{PZ07}
    Portegies Zwart, S.F., Belleman R.G., Geldof P.M., 2007, New Astronomy, 12, 641
%
  \bibitem[\protect\citeauthoryear{Portegies Zwart et al.}{2008}]{PZ08}
    Portegies Zwart, S., McMillan S., Groen D., Gualandris A., Sipior M., Vermin W.,
    2008, New Astronomy, 13, 285
%
\bibitem[Press et al. (2001)]{Press01} Press, W. H., Teukolsky, S. A.,
    Vetterling, W. T., \& Flannery, B. P. 2001, Numerical Recipes
    in Fortran 77 (Cambridge: Cambridge Univ. Press)    
%
  \bibitem[\protect\citeauthoryear{Qian}{1992}]{Q92}
    Qian E., 1992, MNRAS, 257, 581
%    
  \bibitem[\protect\citeauthoryear{Qian}{1993}]{Q93}
    Qian E., 1993, MNRAS, 263, 394    
%  
  \bibitem[\protect\citeauthoryear{Rahmati \& Jalali}{2009}]{RJ09}
    Rahmati A., Jalali M.A., 2009, MNRAS, 393, 1459
%
  \bibitem[\protect\citeauthoryear{Robijn \& Earn}{1996}]{RE96}
    Robijn F.H.A., Earn D.J.D., 1996, MNRAS, 282, 1129
%    
  \bibitem[\protect\citeauthoryear{Saha}{1991}]{Saha91}
    Saha P., 1991, MNRAS, 248, 494   
%    
  \bibitem[\protect\citeauthoryear{Schwarzschild}{1979}]{Sw79}
    Schwarzschild M., 1979, ApJ, 232, 236   
%    
  \bibitem[\protect\citeauthoryear{Springel \& Hernquist}{2002}]{SH02}
    Springel V., Hernquist L., 2002, MNRAS, 333, 649    
%
  \bibitem[\protect\citeauthoryear{Springel}{2005}]{Sp05}
    Springel V., 2005, MNRAS, 364, 1105     
%    
  \bibitem[\protect\citeauthoryear{Sugimoto et al.}{1990}]{S90}
    Sugimoto D., Chikada Y., Makino J., Ito T., Ebisuzaki T.,
    Umemura M., 1990, Nature, 345, 33
%       
  \bibitem[\protect\citeauthoryear{Toomre}{1977}]{T77}
    Toomre A., 1977, ARA\&A, 15, 437 
%
  \bibitem[\protect\citeauthoryear{Weinberg}{1999}]{W99}
    Weinberg M.D., 1999, ApJ, 117, 629
%    
   \bibitem[\protect\citeauthoryear{Zang}{1976}]{Z76}
    Zang T.A., 1976, Ph.D. Thesis, Massachusetts Institute of Technology
%           
   \bibitem[\protect\citeauthoryear{Zhao}{1996}]{Z96}
    Zhao H., 1996, MNRAS, 278, 488
%    
  \bibitem[\protect\citeauthoryear{Zienkiewicz, Taylor \& Zhu}{2005}]{ZT05}
    Zienkiewicz O.C., Taylor R.L., Zhu J.Z., 2005, The Finite Element 
    Method: its basis and fundamentals, 6th edition, 
    Elsevier Butterworth-Heinemann, Oxford
    

\end{thebibliography}
\end{document}